\documentclass[a4paper,11pt]{article}
\pdfoutput=1 
                     
\usepackage[margin=0.8in]{geometry}
\usepackage{latexsym,amssymb,amstext,amsmath}
\usepackage{amsfonts}
\usepackage{enumerate}
\numberwithin{equation}{section}
\usepackage{url}

\usepackage{graphicx}

\usepackage{color}
\usepackage{comment}

\usepackage{amsthm}

\usepackage{esint}

\usepackage{tikz}
\usetikzlibrary{positioning,arrows}
\tikzset{
block/.style={
  draw, 
  rectangle, 
  minimum height=1.5cm, 
  minimum width=4cm, align=center
  }, 
line/.style={->,>=latex'}
}

\usepackage{hyperref}

\def\be {\begin{equation}}
\def\ee {\end{equation}}
\def\bea {\begin{eqnarray}}
\def\eea {\end{eqnarray}}
\def\bal{\begin{align}}
\def\eal{\end{align}}
\def\bsq{\begin{subequations}}
\def\esq{\end{subequations}}

\def\bsp{\begin{split}}
\def\esp{\end{split}}

\begin{document}

\begin{titlepage}

\begin{center}
 {\LARGE\bfseries 
 The gravitational path integral for $ N=4$ BPS black \\[0.3cm] holes from black hole microstate counting }
 \\[10mm]

\textbf{Gabriel Lopes Cardoso$^1$, Abhiram Kidambi$^2$, Suresh Nampuri$^1$, \\ Valentin Reys$^3$, Mart\'i Rossell\'o$^4$}

\vskip 6mm
{\em  $^1$Center for Mathematical Analysis, Geometry and Dynamical Systems,\\
  Department of Mathematics, 
  Instituto Superior T\'ecnico, Universidade de Lisboa,\\
 1049-001 Lisboa, 
 Portugal} \vskip 2mm
 {\em $^2$ Kavli IPMU, University of Tokyo, \\
 Kashiwanoha 5-1-5, 277-8583 Kashiwa, 
 Chiba, Japan} \vskip 2mm
 {\em $^3$ Université Paris-Saclay, CNRS, CEA,\\ Institut de physique théorique, 91191, Gif-sur-Yvette, France.} \vskip 2mm
  {\em $^4$King’s College London, Strand, London WC2R 2LS, United Kingdom}
\vskip 3mm

{\tt 
gabriel.lopes.cardoso@tecnico.ulisboa.pt, abhiram.kidambi@ipmu.jp,
nampuri@gmail.com, valentin.reys@ipht.fr, martirossello@tecnico.ulisboa.pt}\\
\end{center}

\vskip .2in
\begin{center} {\bf ABSTRACT } \end{center}
\begin{quotation}

\noindent 
We use the exact degeneracy formula of single-centred $\frac14$ BPS dyonic black holes
with unit torsion in 4D $N=4$ toroidally compactified heterotic string theory  
to improve on the existing formulation of the corresponding quantum entropy function obtained using supersymmetric localization. The result takes the form of a sum over Euclidean backgrounds including orbifolds of the Euclidean $AdS_2 \times S^2$ attractor geometry. Using an $N=2$ formalism, we determine the explicit form of the Abelian gauge potentials supporting these backgrounds. We further show how a rewriting of the degeneracy formula is amenable, at a semi-classical level, to a gravitational interpretation involving 2D Euclidean wormholes. This alternative picture is useful to elucidate different aspects of the gravitational path integral capturing the microstate degeneracies. We also comment on the relation between the associated 1D holographic models.

\vskip 3mm
\noindent

\end{quotation}
\end{titlepage}

\tableofcontents

\section{Introduction}
\label{sec:intro}
Understanding black hole entropy has served as a fecund nexus of diverse approaches to uncovering the fundamental principles of quantum gravity.
In this note, we analyze some of these approaches and how they feed into each other in the context of 
obtaining the microstate degeneracies $d(m,n,\ell)$ of single-centred 
$\frac{1}{4}$ BPS black holes with unit torsion in 
toroidally compactified 4D $N=4$ heterotic string theory. In this case, there are three distinct approaches to obtaining the corresponding degeneracies for these dyonic configurations:

\begin{enumerate}

\item  The statistical approach, pioneered in \cite{Strominger:1996sh} and used with spectacular success for certain classes of BPS black holes, identifies the statistical microscopic degeneracies as Fourier coefficients of automorphic forms. In the $N=4$ case, 
$d(m,n,\ell)$ are
encoded in the Fourier coefficients of a particular meromorphic Siegel modular form,  namely
the reciprocal of the Igusa cusp form $\Phi_{10}$ \cite{Dijkgraaf:1996it, Dabholkar:2012nd}.
An exact expression for the microstate degeneracies $d(m,n,\ell)$ has been given in \cite{Chowdhury:2019mnb,Cardoso:2021gfg} by resorting to a 
Rademacher type expansion of the Fourier coefficients of $1/\Phi_{10}$, building on earlier work \cite{Bringmann:2010sd,Ferrari:2017msn}. 

\item 

The quantum entropy function approach consists of extracting $d(m,n,\ell)$ from a particular quantum gravity partition function \cite{Sen:2008vm}. In this approach, one is instructed to
sum over gravitational space-time metrics that asymptote to Euclidean 
$AdS_2 \times S^2$.
This {\it a priori} infinite dimensional integral reduces to a finite dimensional one via supersymmetric localization  \cite{Banerjee:2008ky,Banerjee:2009af,Murthy:2009dq,Dabholkar:2014ema}. This drastic simplification enables explicit computations for the counting of BPS black hole microstates. 

\item The holographic approach uses the $AdS_2/CFT_1$ correspondence applied to the decoupled near-horizon geometry associated with asymptotically flat BPS black holes. Bulk fluctuations in $AdS_2$ are holographically encoded in a Lorentzian 1D DFF type model \cite{deAlfaro:1976vlx,Castro:2018ffi,Castro:2019vog,Aniceto:2021xhb}. This model has been shown to capture the semi-classical Wald entropy of BPS black holes \cite{Aniceto:2020saj}.

\end{enumerate}

\begin{figure}
    \centering
    \begin{tikzpicture}

\node[block] (a) {Rademacher \\ expansion};
\node[block, below = 2cm of a] (b) {Quantum entropy\\function};
\node[block, below =2cm of b]   (c){DFF};
\node[block, right =2cm of a]   (d){`Covariant' \\
picture};
\node[block, right =2cm of b]   (e){2D wormholes};
\node[block, right =2cm of c]   (f){1D Liouville};

\draw[line] (d.south) -- (e.north);
\draw[line] (a.south) -- (b.north);
\draw[line] (b.south) -- (c.north);

\draw[line] (e.south) -- (f.north);

\draw[line] (a.east) -- (d.west);
\draw[line] (d.west) -- (a.east);
\draw[line, dashed] (b.east) -- (e.west);
\draw[line, dashed] (e.west) -- node[midway,above] {?} (b.east);

\draw[line] (c.east) -- (f.west);
\draw[line] (f.west) -- (c.east);

\end{tikzpicture}
    \caption{Three approaches to BPS black hole entropy.}
    \label{fig:web}
\end{figure}
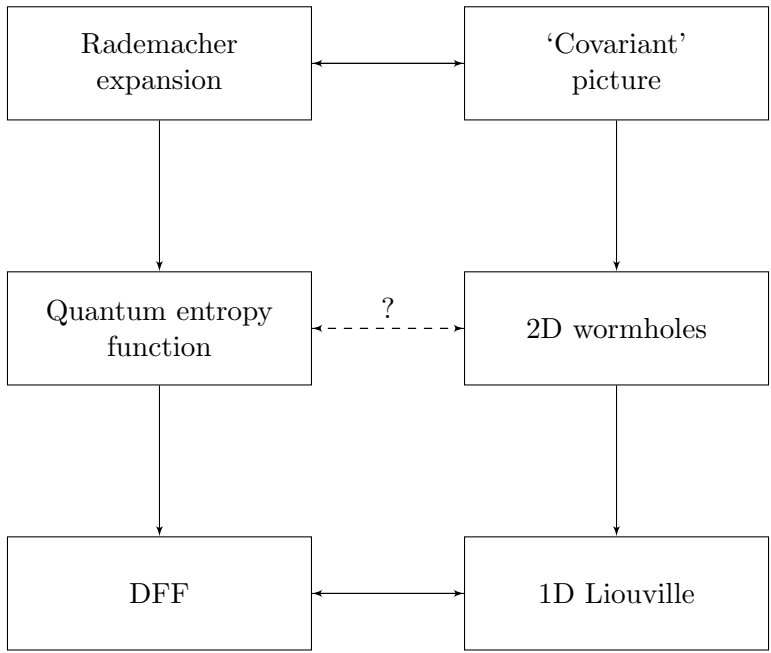

The approaches just discussed are summarized on the left-hand side of Figure \ref{fig:web}. The statistical approach, which is at present the one formulated most rigorously, will be used as a starting point to connect to results obtained in the quantum entropy function approach. We will review the formulation of the quantum entropy function 
in an $N=2$ language. To relate it to the expression for 
$d(m,n,\ell)$ given in \cite{Cardoso:2021gfg} requires certain assumptions that we will spell out in due course. 

Importantly, the expression for 
$d(m,n,\ell)$ given in \cite{Cardoso:2021gfg} can also be brought to a different but equivalent form whose significance is as follows. While the Rademacher form is naturally related to the quantum 
entropy function, which is constructed from an underlying action principle, the equivalent form presents features
that are covariant under S-duality. We will show that this `covariant' picture
suggests a space-time interpretation of the microscopic result in terms of space-time geometries that, at the semi-classical level, involve the 2D Euclidean wormholes in
$AdS_2$ introduced in \cite{Garcia-Garcia:2020ttf,Lin:2022zxd}.

The wormhole interpretation admits a holographic dual description in terms of a 1D Liouville theory \cite{Lin:2022zxd}. We will demonstrate that this model can be rewritten as a Lorentzian DFF type model dual to the near-horizon $AdS_2$ factor by means of an appropriate time reparametrization. This equivalence provides a suggestive first step in understanding the relation between the macroscopic quantum entropy function and 2D wormhole pictures introduced above. The alternative descriptions we propose in this paper, and their interrelations, are summarized on the right-hand side of Figure \ref{fig:web}.

\section{Microstate degeneracies of $\frac14$ BPS $N=4$  black holes}

The degeneracies $d(m,n,\ell)$ of $\frac14$ BPS states with unit torsion in heterotic string theory compactified on a six-torus are given in terms of the Fourier coefficients of 
$1/ \Phi_{10}$, where $\Phi_{10}$ is
the Igusa cusp Siegel modular form of weight $10$ \cite{Dijkgraaf:1996it,Jatkar:2005bh,
David:2006yn}.  $\Phi_{10}$ depends on 3 variables, here denoted by $\rho, \sigma, v$, and the degeneracies
are extracted by integrating  these variables over a contour $C$,
\bea 
	d(m, n, \ell) = (-1)^{\ell+1} \,  \int_C d \sigma  d  v d \rho \,   \frac{1}{\Phi_{10} (\rho, \sigma, v)} \, e^{-2  \pi i \left(  m \rho +  n \sigma + \ell v \right) } \;.
\label{degg}
\eea
Since $ 1/\Phi_{10}$ is a meromorphic Siegel modular form, the resulting expression for the Fourier coefficients $d(m, n, \ell)$ will depend on the choice
of the integration contour $C$.

Here we will focus on the degeneracies of single-centred dyonic $\frac14$ BPS black holes. The microstate degeneracies $d(m,n,\ell)$ of these BPS black holes are determined
in terms of charge bilinears $m, n, \ell$ satisfying 
$\Delta > 0$, where
\bea
\Delta = 4 m n - \ell^2 \;.
\eea
Following \cite{Sen:2007vb,Cheng:2007ch,Sen:2011mh,Dabholkar:2012nd}, 
we define the ${\mathcal R}$-chamber by
\bea
\frac{\rho_2}{\sigma_2} \gg 1 \;,\quad  \frac{v_2}{\sigma_2} = - \frac{\ell}{2m} \;,\quad \frac{\ell}{2m} \in [0, 1) \;,
\label{Rccond}
\eea
where $\rho = \rho_1 + i \rho_2$, $\sigma = \sigma_1 + i \sigma_2$, and $v = v_1 + i v_2$.
To extract  the degeneracies of single-centred $\frac14$ BPS black holes, we take the 
 integration contour $C$ to be the one detailed in \cite{Cardoso:2021gfg}. By first integrating over the variable $\rho$, one obtains an exact
expression for the degeneracies $d(m,n,\ell)$ with $\Delta >0$ in terms of a fine-grained Rademacher type expansion that uses two distinct $SL(2,\mathbb{Z})$ subgroups
of $Sp(4, \mathbb{Z})$. We refer to \cite{Cardoso:2021gfg} for the exact but somewhat lengthy expression for $d(m,n,\ell)$.

When integrating over $\rho$,  one obtains an expression that can be brought to the form \cite{Cardoso:2021gfg}
\begin{equation}
\label{res1}
\begin{split}
d(m,n,\ell)_{\Delta>0} = \sum_P
\sum_{\Sigma \in \mathbb{Z} / |ac| \mathbb{Z}
}\, \frac{(-1)^{\ell+1}}{\gamma^2(ac)^{13}}   \int_{\hat C} \frac{d{\tau} \wedge d{\bar \tau} }{(\tau - {\bar \tau})^{13} }&
\left( \frac{m}{ac} + \frac{a}{c}E_2(\rho'_0)+\frac{c}{a}E_2(\sigma'_0)\right) \times \\
&\quad\frac{1}{\eta^{24}(\rho'_0)}\frac{1}{\eta^{24}(\sigma'_0)}e^{-2\pi i \Lambda} \; ,
\end{split}
\end{equation}
where the set $P$ is defined by
\begin{equation}
	{P} = \left\{ \begin{pmatrix}
		a & b \\
		c&d
	\end{pmatrix}, \begin{pmatrix}
		\alpha & \beta \\
		\gamma & \delta
	\end{pmatrix} \in SL(2,\mathbb{Z})
		 \; \vert \;
	 a,\gamma>0\,, \;c<0\,, \; \alpha, \delta\in\mathbb{Z}/\gamma\mathbb{Z}\,, \; b\in \mathbb{Z}/a\gamma \mathbb{Z} \right\} \;,
	 \label{setP}
\end{equation}
and where $\tau = \tau_1 + i \tau_2$, ${\bar \tau} = \tau_1 - i \tau_2$, and $|\tau|^2 = \tau {\bar \tau}$.
The combinations $\rho'_0$ and $\sigma'_0$ are functions of $\tau$ and $\bar{\tau}$, while the exponent $\Lambda$ depends on $(\tau,\bar{\tau})$ as well as on the charge bilinears $(m,n,\ell)$. Their explicit expressions will be given in the next subsection, where we also discuss the contour $\hat{C}$. The subsequent evaluation of \eqref{res1}
results in the fine-grained Rademacher type expansion mentioned above,
which expresses the BPS degeneracies with $\Delta > 0$ in terms of the polar BPS degeneracies with $\Delta < 0$. We refer to \cite{Cardoso:2021gfg} for more details.

Below we will show that by adding
a total derivative term to the integrand, we can bring the expression \eqref{res1} to the following form, 
\begin{equation}
\label{expvf}
\begin{split}
d(m,n,\ell)_{\Delta>0} = \sum_P
\sum_{\Sigma \in \mathbb{Z} / |ac| \mathbb{Z}
}\,
  \frac{e^{i  \pi \varphi} }{2i\pi} \, \frac{(-1)^{\ell+1} }{\gamma\, (a c )^{13}}  \int_{\hat C} \frac{d \tau \wedge d \bar {\tau}}{(\tau - \bar{\tau})^{14}}&\left( 
26 + \frac{2 \pi}{n_2} \frac{m |\tau|^2 + n - \ell \tau_1}{\tau_2}
\right) \times \\
&\quad\frac{1}{\eta^{24}(\rho'_0)}\frac{1}{\eta^{24}(\sigma'_0)}
\,e^{\frac{\pi}{n_2} \frac{m   | \tau |^2 
+ n  - \ell \tau_1}{\tau_2}  } \; ,
\end{split}
\end{equation}
where $n_2 = -ac\gamma$ and $\varphi$ depends on $(m,n,\ell)$. To write \eqref{expvf}, we have dropped exponentially suppressed contributions associated with small BPS
black holes that we will discuss in Subsection \ref{sec:td}.
As mentioned in the introduction, we will call \eqref{res1} the Rademacher form of the microscopic degeneracy formula, and \eqref{expvf} the `covariant' form.
We will show that the total derivative term that relates the Rademacher and 
`covariant' forms of the degeneracy formula does
not contribute when integrated over the contour $\hat C$.

In the rest of the paper, we  
will explain how both \eqref{res1} and \eqref{expvf} 
are amenable to a gravitational path integral interpretation in terms of a sum over gravitational backgrounds that include Euclidean orbifolds of $AdS_2$ space-times \cite{Banerjee:2008ky,Banerjee:2009af,Murthy:2009dq,Dabholkar:2014ema}.
Furthermore, we will demonstrate that, at a semi-classical level, the `covariant' form \eqref{expvf} is amenable to an alternative macroscopic description in terms of 2D wormholes introduced in \cite{Garcia-Garcia:2020ttf,Lin:2022zxd}.

As shown in \cite{Chowdhury:2019mnb,LopesCardoso:2020pmp}, 
a detailed analysis of the Rademacher form \eqref{res1} reveals that the degeneracies $d(m, n, \ell) $ of $\frac14$ BPS states with $\Delta > 0$ are encoded in the degeneracies of 
polar states, which are BPS states with $\Delta < 0$.  The degeneracies of the latter can be constructed from a continued fraction structure associated with the decay of two-centred $\frac{1}{4}$ BPS bound states. Note that this bound state structure is not manifest in \eqref{expvf}.

In the rest of this section, we first recall how the Rademacher form  \eqref{res1} arises. Subsequently, we show how \eqref{expvf} is related to \eqref{res1}.

\subsection{Integrating $1/ \Phi_{10}$ over $\rho$}
\label{sec:rho-integral}

We first consider integrating \eqref{degg} over $\rho$. To do so we follow \cite{Cardoso:2021gfg}. The integral over $\rho$ is evaluated by summing over the residues in the $\rho$-plane
associated with the quadratic poles of $1/\Phi_{10}$. These poles are characterized by the 5 integers $m_1, n_1, m_2, n_2 \in \mathbb{Z}, \; j \in 2 \mathbb{Z} +1$. One may take $n_2 \geq 0$ without loss of generality. The poles with $n_2>0$ are referred to as quadratic poles, while those with $n_2 =0$
are called linear poles.  The poles 
are defined by the loci
\bea
n_2 (\rho \sigma - v^2) + j v + n_1 \sigma - m_1 \rho + m_2 = 0 \; ,
\label{locpo}
\eea
in the Siegel upper half plane. The five integers $m_1, n_1, m_2, n_2, j$ satisfy the constraint 
\bea
m_1 n_1 + m_2 n_2 = \frac14 \left( 1 - j^2 \right) \;,
\label{mnjrel}
\eea
and they
can be parametrized in terms of 9 integers, 8 of which comprise the entries of the two $SL(2, \mathbb{Z})$ matrices in the set \eqref{setP},
as follows \cite{Murthy:2018bzs,Cardoso:2021gfg} 
\begin{equation}
\label{set1set2enl}
\begin{split}
n_2 =&\; - a c \gamma \;, \\
j =&\; a d + b c \;, \\
n_1 =&\; - b d \alpha - \gamma \Sigma 
\; ,\\
m_1 =&\; a c \delta \; ,\\
m_2 =&\; - b d \beta - \delta  \Sigma  \; .
\end{split}
\end{equation}

Integrating over $\rho$ in this manner, one obtains the expression \cite{Cardoso:2021gfg}
\begin{equation}
\label{afterrho}
\begin{split}
d(m,n,\ell)_{\Delta>0} = \sum_P
\sum_{\Sigma \in \mathbb{Z} / |ac| \mathbb{Z}
}\,
(-1)^{\ell+1}   \int_{\tilde C} d{\sigma} \wedge dv \,\frac{(\gamma\sigma+\delta)^{10}}{ac}&
\left( \frac{m}{ac} + \frac{a}{c}E_2(\rho'_0)+\frac{c}{a}E_2(\sigma'_0)\right)\times \\
&\quad\frac{1}{\eta^{24}(\rho'_0)}\frac{1}{\eta^{24}(\sigma'_0)}e^{-2\pi i \Lambda} \;,
\end{split}
\end{equation}
which is encoded in the two distinct $SL(2,\mathbb{Z})$ subgroups that describe the set $P$ in \eqref{setP}. Here, $E_2$ is the Eisenstein series of weight two and ${\tilde C}$ denotes a particular contour
in the $(\sigma,v)$-plane \cite{Cardoso:2021gfg}.
The variables $\rho'_0$ and $\sigma'_0$ are given in terms of $\sigma$ and $v$ as
\begin{equation}
\label{rsvst}
\begin{split}
\rho'_0 =&\, - \frac{b}{c} \left( \frac{\alpha \sigma + \beta}{\gamma \sigma + \delta}
\right) + \frac{a}{c} \left( \frac{v}{\gamma \sigma + \delta} \right)
  - \frac{a}{c} \Sigma \; ,\\
\sigma'_0 =&\;  \frac{d}{a} \left( \frac{\alpha \sigma + \beta}{\gamma \sigma + \delta}
\right) - \frac{c}{a} \left( \frac{v}{\gamma \sigma + \delta} \right)
  - \frac{c}{a} \Sigma \;,
\end{split}
\end{equation}
while the quantity $\Lambda$ is given by
\begin{equation}
	\Lambda (\sigma, v) = m\left[\frac{\gamma v^2}{\gamma \sigma + \delta} 
-\frac{b d}{ac}   \left( \frac{\alpha \sigma + \beta}{\gamma \sigma + \delta} \right)  + \frac{(a d + b c )}{ac}\; \frac{v}{\gamma \sigma + \delta} - \frac{1}{ac} \Sigma\right] + n\sigma + \ell v\;.
\label{lambb}
\end{equation}

We now bring \eqref{afterrho} to the form \eqref{res1} by expressing 
\bea
 \int_{\tilde C} d{\sigma} \wedge dv \, \frac{(\gamma\sigma+\delta)^{10}}{ac} = -  \int_{{\tilde C}'} \frac{d \rho'_0 \wedge d \sigma'_0 }{\gamma^{13}} 
 \frac{1}{ \left[ \frac{\alpha}{\gamma} - ( c^2 \, \rho'_0 + a^2 \, \sigma'_0 + 2 ac \, \Sigma )\right]^{13}} \;,
 \label{measconv}
 \eea
where the contour ${\tilde C}$ has been mapped to the contour ${\tilde C}'$ through \eqref{rsvst}.
We then obtain for the exponent in \eqref{afterrho},
\bea
\Lambda = \frac{X_1}{X_2} -m  \frac{\Sigma}{ac} \;, 
\label{expXX}
\eea
where
\begin{equation}
\label{X1X2}
\begin{split}
X_1  =&\; \frac{1}{\gamma} \left[ m \left( -  \rho'_0  \sigma'_0  \gamma + \alpha d^2 \rho'_0  + \alpha b^2  \sigma'_0 \right) + n 
\left(  - \beta + \delta A \right) + \ell \left( c d  \rho'_0  + a b  \sigma'_0  \right) 
 \right] \\
 &\; + \Sigma \left( - m  \frac{A}{ac} + m \frac{\alpha}{\gamma} \frac{(a^2 d^2 + b^2 c^2 )}{ac} + 2 \, n \,  \frac{\delta}{\gamma} a c + \ell \,
 \frac{ad + bc}{\gamma} 
 \right) - \Sigma^2 \, m \;,\\
X_2 =&\; \frac{\alpha}{\gamma} - A - 2  a c \Sigma \;,
\end{split}
\end{equation}
and where for later use we define the combinations
\begin{equation}
\label{AB} 
\begin{split}
A =&\;  c^2 \, \rho'_0 + a^2 \, \sigma'_0 \; , \\
B =&\; c^2 \, \rho'_0 - a^2 \, \sigma'_0 \; ,
\end{split}
\end{equation}
in terms of which
\bea
\rho'_0 = \frac{A + B}{2 c^2} \;,\quad \sigma'_0  =  \frac{A - B}{2 a^2} \; .
\label{rsAB}
\eea

Next we change integration variables from $(\rho'_0, \sigma'_0)$ to $(\tau, {\bar \tau})$ using 
\begin{equation}
\label{newv}
\begin{split}
\rho'_0  =&\, - \frac{a}{c} \, \frac{\tau}{\gamma}  - \frac{b}{c} \, \frac{\alpha}{\gamma} - \frac{a}{c} \Sigma \;, \\
\sigma'_0 =&\; \frac{c}{a} \frac{{\bar \tau}}{\gamma}  + \frac{d}{a} \, \frac{\alpha}{\gamma}  - \frac{c}{a} \Sigma \;,
\end{split}
\end{equation}
where we define 
$\tau = \tau_1 + i \tau_2$, ${\bar \tau} = \tau_1 - i \tau_2$, and $|\tau|^2 = \tau {\bar \tau}$. Note that $\tau_1$ and $\tau_2$ are complex variables, since 
$\rho'_0$ and $\sigma'_0 $ are complex.
The quantities $X_1$ and $X_2$ are expressed in terms of $\tau$ and $\bar{\tau}$ as
\begin{equation}
\label{x1x2t}
\begin{split}
X_2 =&\; \frac{ac}{\gamma} \left( \tau - {\bar \tau} \right) \,=\, 2 i  \frac{ac}{\gamma}  \tau_2 \; ,\\
\gamma X_1 =&\; m \left( \frac{ \tau \bar \tau}{\gamma} - \frac{\alpha}{\gamma} b d ( \tau - {\bar \tau}) \right)
+ n \left( \frac{1}{\gamma} - \frac{\delta}{\gamma} a c ( \tau - {\bar \tau}) \right) - \ell \frac{\tau_1}{\gamma} - i \ell \frac{\tau_2}{\gamma} (ad + bc) \;, 
\end{split}
\end{equation}
and hence we obtain for the combination \eqref{expXX},
\bea
\label{Lsvmn}
-2\pi i \Lambda
= \frac{\pi}{n_2}  \left[
\frac{ m  |\tau|^2 + n - \ell \tau_1}{\tau_2} + 2 i \left( - \frac12 j \ell - m_1 n + n_1 m \right) 
\right].
\eea
Then, \eqref{afterrho} becomes 
\begin{equation}
\label{res}
\begin{split} 
d(m,n,\ell)_{\Delta>0} = \sum_P
\sum_{\Sigma \in \mathbb{Z} / |ac| \mathbb{Z}
}\,
\frac{(-1)^{\ell+1}}{\gamma^2 \, (ac)^{13}}   \int_{\hat C} \frac{d{\tau} \wedge d{\bar \tau} }{(\tau - {\bar \tau})^{13} }
&\left( \frac{m}{ac} + \frac{a}{c}E_2(\rho'_0)+\frac{c}{a}E_2(\sigma'_0)\right) \times \\
&\quad\frac{1}{\eta^{24}(\rho'_0)}\frac{1}{\eta^{24}(\sigma'_0)}e^{-2\pi i \Lambda}\;,
\end{split}
\end{equation}
with $\rho'_0$, $\sigma'_0$ and $\Lambda$ expressed in terms of $\tau$ and $\bar{\tau}$ through
\eqref{newv} and \eqref{Lsvmn}, and where the contour ${\tilde C}'$ has been mapped to the contour ${\hat C}$ through \eqref{newv}. 
To describe the new contour 
${\hat C}$, we first note the relations
\begin{equation}
\begin{split} 
\tau =&\; - \frac{\gamma (v + \frac{b}{a \gamma})}{\gamma \sigma + \delta}
= \tau_1 + i \tau_2 \;, \\
{\bar \tau} =&\;
- \frac{\gamma (v + \frac{d}{c \gamma})}{\gamma \sigma + \delta}
= \tau_1 - i \tau_2 \;.
\end{split}
\end{equation}
Then, using the relation \eqref{Rccond} and the contours
of integration over $\sigma$ and $v$ discussed in \cite{Cardoso:2021gfg}, it can be
shown that ${\hat C}$
takes the following form
in terms of the $(\tau_1,\tau_2)$ coordinates,
\begin{equation}
\label{t12c}
    \tau_1 \;: \; \frac{\ell}{2m}-i\tau_2 \; \to \; \frac{\ell}{2m}+i\tau_2\;,\quad \tau_2 \;: \; \varepsilon-i\infty \; \to \; \varepsilon+i\infty\;,
\end{equation}
where $\varepsilon$ is a positive real number. We can set it to $\varepsilon =\frac{\sqrt{\Delta}}{2m}$ to guarantee that the contour passes through the attractor point, which is the  tree-level saddle point value for $\tau$ given below in \eqref{tau}. The contour \eqref{t12c} can be parametrized by
\begin{equation} 
\label{g1g2}
\begin{split}
\Gamma_1\;: \;\tau_1 =&\; \frac{\ell}{2m} + i \tau_2 \left( - 1 + 2 y \right)
\;, \quad \delta \leq y \leq 1 - \delta \; , \\
\Gamma_2\;: \;\tau_2 =&\; \varepsilon + i t \;,\quad -\infty < t < \infty \; ,
\end{split}
\end{equation}
where $\delta > 0$ is a  regulator, whose presence ensures
that the arguments of the $\eta$ functions in \eqref{res}
satisfy $| e^{2 \pi i \rho'_0} | < 1$ and 
$| e^{2 \pi i \sigma'_0} | < 1$ on the contour.\footnote{A similar contour was also used in \cite{Gomes:2015xcf,Murthy:2015zzy}.}
Indeed, on this contour we have
\begin{equation}
\begin{split}
\tau_1 + i \tau_2 =&\; \frac{\ell}{2m} + 2 i \tau_2 y = \frac{\ell}{2m} - 2 t y + 2 i \varepsilon y \; , \\
\tau_1 - i \tau_2 =&\; \frac{\ell}{2m} - 2 i \tau_2 (1-y) = \frac{\ell}{2m} + 2 t (1-y)  - 2 i  \varepsilon (1-y) \; ,
\end{split}
\end{equation}
and hence, using \eqref{newv}, we infer that on the contour
\begin{equation}
\begin{split}
| e^{2 \pi i \rho'_0 } | =&\; e^{  4 \pi \frac{a  }{c \gamma}  \varepsilon y  } = e^{  - 4 \pi \frac{a  }{|c| \gamma}  \varepsilon y  } < 1 \;, \\
| e^{2 \pi i \sigma'_0 } | =&\; e^{  4 \pi \frac{c  }{a \gamma}  \varepsilon (1-y)  } = e^{  - 4 \pi \frac{|c|  }{a \gamma}  \varepsilon (1-y)  } < 1 \;.
\end{split}
\end{equation}

Throughout the paper, when we write $\int_{\hat C} d{\tau} \wedge d{\bar \tau}  \, f$, we will mean
\bea
2 i \int_{\Gamma_2} d \tau_2 \left( \int_{\Gamma_1} d\tau_1 \, f \right) .
\eea
In the next subsection we will show how the Rademacher form \eqref{res} of the degeneracy formula can be brought into the 
`covariant' form \eqref{expvf}.

\subsection{Adding a total derivative term \label{sec:td}}

The Rademacher form \eqref{res} is obtained by performing the integration over the $\rho$ variable in \eqref{degg}. An alternative expression for the degeneracies can be obtained by first integrating over the $v$ variable. As shown in \cite{Banerjee:2008ky}, the leading contribution to \eqref{degg} coming from the residue at the pole $\rho\sigma - v^2 + v=0$ takes the form of the integral entering \eqref{expvf}. We now explain how to directly relate the degeneracies obtained from integrating over $\rho$ and over $v$ by the addition of a total derivative term.

We start from \eqref{res}, 
\begin{equation}
\label{resreg}
\begin{split} 
d(m,n,\ell)_{\Delta>0} = \sum_P
\sum_{\Sigma \in \mathbb{Z} / |ac| \mathbb{Z}
}\,  e^{i  \pi \varphi}  \, 
\frac{(-1)^{\ell+1}}{\gamma^2 \, (ac)^{13}}   \int_{\hat C} \frac{d{\tau} \wedge d{\bar \tau} }{(\tau - {\bar \tau})^{13} }
&\left( \frac{m}{ac} + \frac{a}{c}E_2(\rho'_0)+\frac{c}{a}E_2(\sigma'_0)\right) \times \\
&\quad\frac{1}{\eta^{24}(\rho'_0)}\frac{1}{\eta^{24}(\sigma'_0)}\,
e^{ \frac{\pi}{n_2}
\frac{ m  |\tau|^2 + n - \ell \tau_1}{\tau_2} 
}
\;,
\end{split}
\end{equation}
where the phase $\varphi$ reads
\bea
\varphi = 
  \frac{2}{n_2} \left( - \frac12 j \, \ell - m_1 \, n + n_1 \, m
  \right) \, .
  \label{phg1}
  \eea
To this we add the total derivative term
\bea
\label{tdt}
\int_{\Gamma_2} d\tau_2\, \frac{d}{d \tau_2} \left( \int_{\Gamma_1} d\tau_1 \, R(\tau_1, \tau_2) \right) \;,
\eea
with
\bea
R(\tau_1, \tau_2) = \frac{i}{\pi} 
\sum_P
\sum_{\Sigma \in \mathbb{Z} / |ac| \mathbb{Z}
}\,  e^{i  \pi \varphi}  \, 
\frac{(-1)^{\ell+1}}{\gamma \, (2 i ac)^{13}}
\, \frac{1}{\tau_2^{13}} \frac{1}{\eta^{24} ( \rho'_0 )}  \frac{1}{\eta^{24} ( \sigma'_0 )} \, e^{ \frac{\pi}{n_2}
\frac{ m  |\tau|^2 + n - \ell \tau_1}{\tau_2} 
}
\;,
\eea
to obtain
\begin{equation}
\label{covqef}
\begin{split}
d(m,n,\ell)_{\Delta>0} =&\; \sum_P
\sum_{\Sigma \in \mathbb{Z} / |ac| \mathbb{Z} 
}\Biggl\{\,\frac{e^{i  \pi \varphi} }{ 2 i\pi } \, \frac{(-1)^{\ell +1} }{\gamma\, (a c )^{13}} \int_{\hat C} \frac{d \tau \wedge d \bar {\tau}}{(\tau - \bar{\tau})^{14}}  \left( 
26  + \frac{2\pi}{n_2} \,
\frac{ m  |\tau|^2 + n - \ell \tau_1}{\tau_2}  
\right)
\times\\
&\hspace{70mm}\frac{1}{\eta^{24}(\rho'_0)}\frac{1}{\eta^{24}(\sigma'_0)} \, 
e^{ \frac{\pi}{n_2}
\frac{ m  |\tau|^2 + n - \ell \tau_1}{\tau_2} 
}\Biggr\} 
\\
&\, +  i  (1 - 2\delta)  \int_{\Gamma_2} \, d \tau_2 \, R \left(\tau_1 = \frac{\ell}{2m} \!\!+ i \tau_2 (1 - 2\delta), \tau_2 \right)
\\
&\,
+ i  (1 - 2\delta)  \int_{\Gamma_2} \, d \tau_2 \, R \left(\tau_1 = \frac{\ell}{2m} \!\!- i \tau_2 (1 - 2\delta), \tau_2 \right) \; .
\end{split}
\end{equation}

Next, let us analyze the 
last two terms in \eqref{covqef}.
We focus on the term 
\begin{equation}
\label{R1}
\begin{split}
\int_{\Gamma_2} d \tau_2 \, 
R \left(\tau_1 = \frac{\ell}{2m} \!\!+ i \tau_2 (1 - 2\delta), \tau_2 \right) =&\; \frac{i}{ \pi} 
\sum_P
\sum_{\Sigma \in \mathbb{Z} / |ac| \mathbb{Z}
}\,  e^{i  \pi \varphi}  \, 
\frac{(-1)^{\ell+1}}{\gamma \, (2 i ac)^{13}} \, \\
&\quad\int_{\Gamma_2} \frac{d \tau_2}{\tau_2^{13}} \, 
\,  \frac{1}{\eta^{24} ( \rho'_0 )}  \frac{1}{\eta^{24} ( \sigma'_0 )} \, 
e^{ \frac{\pi}{n_2}  \left[
\frac{ \Delta}{4m \tau_2} + 4  m \tau_2 \, \delta (1 - \delta) \right]} 
\;,
\end{split}
\end{equation}
where
\begin{equation}
\begin{split}
\rho'_0  =&\; -  \frac{a  }{c \gamma} \, \left(  \frac{\ell}{2m} + 2 i \tau_2 (1 - \delta)  \right) - \frac{1}{c} \left( b \frac{\alpha}{\gamma} + a \Sigma \right) \; ,\\
\sigma'_0 =&\; \frac{c  }{a \gamma} \, \left( \frac{\ell}{2m} -2 i \tau_2 \delta  \right) + \frac{1}{a} \left( d \frac{\alpha}{\gamma} - c  \Sigma \right) \; .
\end{split}
\end{equation}
Let us perform a saddle point analysis of the integral in \eqref{R1}. Taking ${\rm Im}\,  \tau_2$ to be large,
and using the approximation
\bea
\frac{1}{\eta^{24}(\rho'_0)}\frac{1}{\eta^{24}(\sigma'_0)} \approx  e^{- 2 \pi i \left( \rho'_0 +  \sigma'_0 \right)},
\eea
we obtain for the saddle point value of $\tau_2$,
\bea
\tau_{2 \, \text{extr}}  = \frac{\sqrt{\Delta}}{
\sqrt{16m\, (|a|^2 (1- \delta) + |c|^2 \delta + m \delta (1 - \delta) } }
\;.
\eea
For the case $n_2 = 1$, and setting $\delta =0$, this becomes
\bea
\tau_{2 \, \text{extr}}  = \sqrt{ \frac{\Delta}{16m}} \;,
\eea
which, when inserted back into \eqref{R1}, shows that the summand on the right-hand side behaves as $e^{2 \pi \sqrt{\Delta/m}}$ in a saddle point approximation. This is an exponentially supressed contribution that can be dropped in the semi-classical expansion, as it does not scale with the charges. A similar argument applies to the last two terms of \eqref{covqef}. 
Since in Section~\ref{sec:sti} we will be interested in the semi-classical analysis of dyonic single-centred BPS black holes whose classical entropy scales quadratically in the charges,
we will ignore these terms in what follows.

Finally, let us show that the integrated total derivative term \eqref{tdt} vanishes,
\bea
\int_{\Gamma_2}  \frac{d}{d \tau_2} \left( \int_{\Gamma_1} d\tau_1 \, R(\tau_1, \tau_2) \right) 
= \frac{i}{ \pi} 
\sum_P
\sum_{\Sigma \in \mathbb{Z} / |ac| \mathbb{Z}
}\,  e^{i  \pi \varphi}  \, 
\frac{(-1)^{\ell+1}}{\gamma \, (2 i ac)^{13}} \left( 
G_{n_2}(\varepsilon + i \infty) - G_{n_2} (\varepsilon - i \infty) \right) , 
\eea
where 
\bea
G_{n_2}(\tau_2) 
= \frac{2i}{\tau_2^{12} } \, e^{ \frac{ \pi\Delta}{ 4 m n_2 \tau_2} }
 \int_{\delta}^{1 - \delta} dy \, \frac{1}{\eta^{24} ( \rho'_0 )}  \frac{1}{\eta^{24} ( \sigma'_0 )} \, e^{ \frac{4 \pi m \tau_2 y (1-y)}{n_2} } .
\eea
Let us estimate $|G_{n_2}(\tau_2)|$ using Darboux's inequality,
\bea
|G_{n_2} (\varepsilon + i t) | \leq \frac{2}{
(\varepsilon^2 + t^2)^{6}} \, e^{ \frac{ \pi\Delta \varepsilon}{ 4 m n_2\sqrt{ \varepsilon^2 + t^2}} }
\, e^{ \frac{ \pi m \varepsilon }{n_2} } \, M (t)  \;,\quad M (t) = 
\int_{\delta}^{1 - \delta} dy \, { \Bigg \vert}  \frac{1}{\eta^{24} ( \rho'_0 )}  \frac{1}{\eta^{24} ( \sigma'_0 )} { \Bigg \vert} \; .
\eea
Note that $M(t)$ is finite due to the regulator $\delta$, and that it stays finite for arbitrary values of $t$, i.e. $M(t) \leq M \;\; \forall t \in \mathbb{R}$.  Therefore $|G_{n_2} (\varepsilon \pm i \infty ) | =0$, and hence  $G_{n_2} (\varepsilon \pm i \infty )  =0$. Thus, we have established the equivalence of the Rademacher and `covariant' forms 
\eqref{res1} and \eqref{expvf}, respectively, up to exponentially suppressed corrections. In Section \ref{wormholes}, we will give a physical motivation for adding the total derivative term to \eqref{res} and working with the covariant expression of the degeneracies, which allows for an interpretation involving 2D Euclidean wormholes.

Note that in the `covariant' expression \eqref{expvf} the measure factor $26 +  \frac{2 \pi}{n_2} \frac{m |\tau|^2 + n - \ell \tau_1 }{\tau_2}$ does not carry any dependence on $\Sigma$,  and that the dependence of the integrand on the $SL(2, \mathbb{Z}) $ parameters of \eqref{setP} is solely contained in the combination $n_2 = - a c \gamma$, cf. \eqref{set1set2enl}.

\subsection{Semi-classical expansions \label{sec:sti} }

We established that the expression for the microstate degeneracies \eqref{resreg} is equal to the expression \eqref{expvf} (up to the exponentially suppressed contributions just discussed) 
which we reproduce here for convenience, 
\begin{equation}
\label{covqef2}
\begin{split}
d(m,n,\ell)_{\Delta>0} = \sum_P
\sum_{\Sigma \in \mathbb{Z} / |ac| \mathbb{Z}
}\,
  \frac{e^{i  \pi \varphi} }{2i\pi} \, \frac{(-1)^{\ell+1} }{\gamma\, (a c )^{13}}  \int_{\hat C} \frac{d \tau \wedge d \bar {\tau}}{(\tau - \bar{\tau})^{14}}&\left( 
26 + \frac{2 \pi}{n_2} \frac{m |\tau|^2 + n - \ell \tau_1}{\tau_2}
\right) \times \\
&\quad\frac{1}{\eta^{24}(\rho'_0)}\frac{1}{\eta^{24}(\sigma'_0)}
\,e^{\frac{\pi}{n_2} \frac{m   | \tau |^2 
+ n  - \ell \tau_1}{\tau_2}  } \; ,
\end{split}    
\end{equation}
where
\begin{equation} 
\label{rstexp}
\begin{split}
\rho'_0  =&\, - \frac{a}{c} \, \frac{\tau}{\gamma}  - \frac{b}{c} \, \frac{\alpha}{\gamma} - \frac{a}{c} \Sigma \;, \\
\sigma'_0 =&\; \frac{c}{a} \frac{{\bar \tau}}{\gamma}  + \frac{d}{a} \, \frac{\alpha}{\gamma}  - \frac{c}{a} \Sigma \; .
\end{split}
\end{equation}
Let us consider the integrand of \eqref{covqef2}. Both the measure and the exponent are given in terms of the combination
\bea
\frac{ \pi}{n_2} \frac{m |\tau|^2 + n - \ell \tau_1}{\tau_2} \;,
\label{treelcomb} 
\eea
which, upon substituting the tree-level saddle point value for $\tau$,
is equal to 
\bea
\frac{{\cal S}_{\rm BH}}{n_2} = \frac{\pi \sqrt{\Delta}}{n_2}\;,
\eea
where ${\cal S}_{\rm BH}$ denotes the Bekenstein-Hawking entropy of a BPS black hole carrying charge bilinears $(m, n, \ell)$ at the two-derivative level.
Note that despite the presence of the combination \eqref{treelcomb} in the measure of \eqref{covqef2},
no contribution proportional to $\log {\cal S}_{\rm BH}$ arises
when evaluating the integral \eqref{covqef2} in saddle point approximation \cite{Banerjee:2008ky}. 
When $n_2 =1$, the associated quantum entropy function space-time background is $AdS_2 \times S^2$, while when $n_2 >1$, the associated space-time background has been argued to be a 
freely acting $\mathbb{Z}_{n_2}$ orbifold of an $AdS_2 \times S^2$ background
\cite{Banerjee:2008ky,Banerjee:2009af,Murthy:2009dq,Dabholkar:2014ema}.
We note at this stage that the order $n_2$ of the orbifold is a composite integer given by
$n_2 = - a c \gamma$. Moreover, the orbifold
background is supported by electric and magnetic gauge potentials $(A_{\theta}^I, {\tilde A}_{\theta I})$ whose values depend on the integers introduced in \eqref{set1set2enl}. There are therefore several sectors in \eqref{covqef2} that are 
labelled by different integers but have the same order $n_2$. 
We will discuss these 
orbifold backgrounds in more detail in Section \ref{sec:orbi}, in conjunction with S-duality.\\

To facilitate the discussion about the space-time interpretation of \eqref{resreg} and \eqref{covqef2}, it is useful
to first infer semi-classical expressions for $\log d(m, n, \ell)$ by extremizing the integrand of either expressions. Both extremizations give identical results, and
we will choose to extremize the latter since the analysis is less involved.
We will first extremize the terms in the exponent of \eqref{covqef2}, and subsequently consider the extremization problem in the presence
of the $\eta^{24}$ terms.

For subsequent use, we recall that the evaluation of \eqref{resreg} results in an exact expression for the degeneracies
$d(m, n, \ell)$ in terms of a fine-grained Rademacher expansion that involves Kloosterman sums and modified Bessel functions of the first 
kind \cite{Cardoso:2021gfg}. This exact expression, given in equation (5.118) of \cite{Cardoso:2021gfg}, is rather lengthy and will not be reproduced here. For our semi-classical analysis it will be sufficient to focus on the first line of this result since, in the large $\Delta$ limit, the remaining terms are exponentially supressed. Here we simply recall that this part of the exact degeneracies involves modified Bessel functions $I_{23/2} (z)$, with 
$z = \pi \sqrt{\Delta | {\tilde \Delta} | }/(\gamma m)$, as well as
the generalized Kloosterman sum
\begin{equation}
\label{kloosterms}
\begin{split}
{\rm Kl}\Bigl(\frac{\Delta}{4m},\frac{\tilde{\Delta}}{4m},\gamma,\psi\Bigr)_{\ell\tilde{\ell}} =&\; 
 \sum_{\substack{0\leq -\delta <\gamma\\ (\delta,\gamma)=1,\, \alpha\delta = 1 \text{ mod } \gamma}}e^{2\pi i \left( \frac{\alpha}{\gamma}\frac{\tilde{\Delta}}{4m} +\frac{\delta}{\gamma}\frac{\Delta}{4m}\right)} \, {\tilde \psi} (\Gamma)_{\tilde{\ell}\ell} \;,\\
	 {\tilde \psi} (\Gamma)_{{\tilde \ell} \ell} =&\; - \frac{1}{\sqrt{2m\gamma }}\sum_{T\in\mathbb{Z}/\gamma\mathbb{Z}} e^{2\pi i \left(\frac{\alpha}{\gamma}\frac{({\tilde \ell }-2mT)^2}{4m}-\frac{\ell ({\tilde \ell} -2mT)}{2m\gamma} +\frac{\delta}{\gamma}\frac{\ell^2}{4m}   \right)} \;,
\end{split}
\end{equation}
where 
\bea
\Delta = 4 m n - \ell^2 > 0 \;,\quad {\tilde \Delta} = 4 m {\tilde n} - {\tilde \ell}^2 < 0\;.
\label{ddt}
\eea

\subsubsection{Extremization without $\log \eta$ terms}

Let us consider the terms in the exponent of \eqref{covqef2},
\bea
f_0(\tau,\bar{\tau}) = \frac{\pi}{n_2} \frac{ m   | \tau |^2 
+ n  - \ell \tau_1 }{\tau_2}  + i \pi  \varphi \; , 
\label{expc}
\eea
where we recall that the phase $\varphi$ is given by \eqref{phg1}.
Extremizing  with respect to $\tau$ yields
\bea
\tau_* =  \frac{\ell}{2 m} + i \frac{\sqrt{\Delta}}{2 m} \; .
\label{tau}
\eea
At this extremum, \eqref{expc} takes the value
\bea
f_0(\tau_*,\bar{\tau}_*) = \pi \left[  \frac{\sqrt{\Delta}}{n_2}- \frac{i}{n_2} (ad + bc ) \ell  + 2 i \frac{\delta}{\gamma} n +2  i \frac{\alpha}{\gamma}
\frac{b d}{ac} \, m + 2  i m \frac{\Sigma}{ac} \right]\;.
\label{expmn2} 
\eea
which we can write as 
\bea
f_0(\tau_*,\bar{\tau}_*) = \pi \left[  \frac{\sqrt{\Delta}}{n_2} + \frac{2 i}{n_2} \left( - \frac12  j \ell - m_1 n + n_1  m \right)
\right],
\label{expmn} 
\eea
using \eqref{set1set2enl}. We thus reproduce the semi-classical result for $\log d(m, n, \ell)$ obtained in
 \cite{Sen:2007qy,Banerjee:2008ky, Murthy:2009dq} as expected.

Next, we determine the values of $(\rho, \sigma, v)$ at the extremum, which we will denote by
$(\rho_*, \sigma_*, v_*)$. Using
\begin{equation}
\label{exprsv}
\begin{split}
\rho =&\; \frac{\frac{\alpha}{\gamma} \left( d^2 \, \rho'_0 + b^2 \,  \sigma'_0 + 2 b d \Sigma  \right) +  \Sigma^2 - \rho'_0  \sigma'_0 }{X_2} \;, \\
\sigma =&\;  - \frac{\delta}{\gamma} + \frac{1}{\gamma^2 \, X_2 } \;, \\
v =&\; \frac{ c d \rho'_0 + a b  \sigma'_0+ (a d + b c ) \Sigma  }{\gamma \, X_2} \;, \\
\gamma \sigma + \delta =&\; \frac{1}{\gamma \, X_2} \;,
\end{split}
\end{equation}
as well as \eqref{newv} and \eqref{tau},
we obtain
\begin{equation}
\label{saddr}
\begin{split}
\rho_* =&\; - \frac{n_1}{n_2} + i\frac{n}{n_2\sqrt{\Delta}} \; , \\
\sigma_* =&\; \frac{m_1}{n_2} + i\frac{m}{n_2\sqrt{\Delta}} \; , \\
v_* =&\; \frac{j}{2 n_2} - i\frac{\ell}{n_2\sqrt{\Delta} } \; .
\end{split}
\end{equation}
This reproduces the result for $(\rho_*, \sigma_*, v_*)$ obtained in \cite{Murthy:2009dq}, apart from a sign difference in ${\rm Im} \, v_*$.

\subsubsection{Extremization with $\log \eta$ terms \label{sec:ext-redone}}

Next, we consider the extremization of the terms in the exponent in \eqref{covqef2}, but this time
also taking into account the $\eta^{24}$ terms in \eqref{covqef2}. This will result in a change of the saddle point value of
$\tau $ given in \eqref{tau}.

We take the imaginary part of $\tau$ to be large and use the approximation
\bea
\frac{1}{\eta^{24}(\rho'_0)}\frac{1}{\eta^{24}(\sigma'_0)} \approx  e^{- 2 \pi i \left( \rho'_0 +  \sigma'_0 \right)} .
\label{etetapp}
\eea
We thus extremize the function
\bea
f(\tau,\bar{\tau}) = - 2 \pi  i \left( \rho'_0 +  \sigma'_0 \right)  + \pi \frac{ m   | \tau |^2 
+ n  - \ell \tau_1 }{n_2 \,  \tau_2} + i \pi \varphi
\label{eeps}
\eea
with respect to $\tau$ and $\bar \tau$ using the expressions \eqref{newv}. This results in the following two equations,
\begin{equation}
\begin{split}
a^2 (\tau - {\bar \tau})^2 + m {\bar \tau}^2 - \ell {\bar \tau} + n =&\; 0 \;, \\
c^2 (\tau - {\bar \tau})^2 + m \tau^2 - \ell  \tau + n =&\; 0 \;, 
\end{split}
\end{equation}
which are solved by
\begin{equation}
\begin{split}
\tau_{\text{extr}} =&\;   \frac{ \ell}{2 m} +  i \frac{\sqrt{\Delta}}{2 m}  \frac{m + a^2 - c^2   }{\sqrt{ (m +  (a-c)^2) (m +  (a + c)^2) }}
\;, \\
{\bar \tau}_{\text{extr}} =&\;   \frac{ \ell}{2 m} - i 
 \frac{\sqrt{\Delta}}{2 m}  \frac{m - a^2 + c^2   }{\sqrt{ (m +  (a-c)^2) (m +  (a + c)^2) }} ,
\label{act}
\end{split}
\end{equation}
where we chose the appropriate signs so that \eqref{act} reduces to \eqref{tau} when switching off the terms proportional to $a^2$ and $c^2$.
Note that when $a^2 \neq c^2$,  the saddle point values $\tau_{\text{extr}}$ and ${\bar \tau}_{\text{extr}}$ are not any longer complex conjugates of one another.

In the following, let us consider the case $a = - c =1, b=0, d=1$. The
saddle point value is then
\bea
\tau_{\text{extr}} = 
  \frac{ \ell}{2 m} + i \frac{\sqrt{\Delta}}{2 \sqrt{(m +  4) m }}  \;,
\eea
which reproduces the $R^2$-corrected attractor value for $\tau$ at leading order \cite{LopesCardoso:1999fsj}.
Next we 
evaluate the value of \eqref{eeps} at the saddle point. Exponentiating the resulting expression gives
\bea
e^{f(\tau_{\text{extr}},\bar{\tau}_{\text{extr}})} = e^{ 2 \pi i \left(  n \frac{\delta}{\gamma} -  \frac{\alpha}{\gamma}  - \frac{\ell}{2 \gamma} \right) + 
\frac{\pi}{\gamma}  \sqrt{\Delta} \frac{ \sqrt{m + 4}}{\sqrt{m}} } .
\label{r2corrph}
\eea
Comparing with \eqref{expmn2}, we note the presence of the additional $R^2$ induced phase $\alpha$ stemming from the terms \eqref{etetapp}.

We observe that the exponent in \eqref{r2corrph} corresponds to the term 
${\tilde n} = -1, {\tilde \ell} = m$ in the Rademacher expression given in (5.118) of \cite{Cardoso:2021gfg}, in which case 
${\tilde \Delta}  = 4 m {\tilde n} -  {\tilde \ell}^2 = - m (4 +m)$ .
The real part of the exponent yields Wald's $R^2$ corrected entropy \cite{LopesCardoso:1999fsj} divided
by $\gamma$.
The imaginary part of the exponent matches the exponent in the Kloosterman sum \eqref{kloosterms}
\begin{equation}
\label{kloostexp}
\begin{split}
\frac{\delta}{\gamma} \frac{\Delta}{4m} + \frac{\alpha}{\gamma} \frac{{\tilde \Delta}}{4m} + \frac{\alpha}{\gamma} \frac{( {\tilde \ell} - 2 m T)^2}{4m} 
&- \frac{ \ell ( {\tilde \ell} - 2 m T)}{2 m \gamma} + \frac{\delta}{\gamma} \frac{\ell^2}{4m} \\
=&\;\frac{\delta}{\gamma} n + \frac{\alpha}{\gamma} \left( {\tilde n} - {\tilde \ell} T + m T^2 \right) - \frac{ \ell ( {\tilde \ell} - 2 m T)}{2 m \gamma} \; ,
\end{split}
\end{equation}
for
\bea
{\tilde n} = -1 \;,\quad {\tilde \ell} = m \;,\quad T = 0 \; .
\eea
Note that the $R^2$ induced phase $\alpha$ is crucial to obtain this matching.

As mentioned above, the exact expression for the microstate degeneracies $d(m,n,\ell)$ involves modified
Bessel functions $I_{23/2}(z)$, with $z = \pi \sqrt{\Delta | {\tilde \Delta} | }/(\gamma m)$.
Note that $z$ has a dependency on $1/\gamma$, not on $1/n_2$. On the other hand, the 
saddle point result \eqref{expmn}, when exponentiated, gives an expression whose modulus is
$e^{\pi \sqrt{\Delta}/n_2}$. One may then wonder how these two seemingly differently looking expressions are
related. To do so, 
we proceed as follows. The quantity $\tilde \Delta$ in \eqref{ddt} can be expressed in terms of integers $M,N$ as \cite{Cardoso:2021gfg}
\bea
{\tilde \Delta} = \frac{1}{a^2 c^2} \left[ - (a^2 M - c^2 N)^2 - (m^2 - 2m (a^2 M + c^2 N)) \right],
\eea
where $M,N$ satisfy the bounds
\bea
- m < a^2 M - c^2 N \leq m \;,\quad a^2 M + c^2 N \leq  m \;.
\eea
Taking $m$ to be large, and restricting to integers $M,N$ such that the following holds, 
\bea
| a^2 M - c^2 N| \ll m \;,\quad |a^2 M + c^2 N| \ll m \;,
\label{rangevalMN}
\eea
we find that 
${\tilde \Delta}$ is approximately given by
\bea
|{\tilde \Delta}| \approx \frac{m^2}{(a c )^2} \;,
\eea
in which case $z$ becomes 
\bea
z = \frac{\pi \sqrt{\Delta}}{n_2} \;,
\eea
so that to leading order, 
\bea
I_{23/2} (z) \approx e^{ \frac{\pi \sqrt{\Delta}}{n_2} } \;.
\label{In2}
\eea
Note that the set of integers  $M,N$ satisfying \eqref{rangevalMN} is finite. This is a consequence
of demanding that ${\tilde \Delta} < 0$ as well as of the continued fraction condition stated in \cite{Cardoso:2021gfg}.

\section{Orbifolds of $AdS_2$  and S-duality \label{sec:orbi} }

The microscopic result \eqref{covqef2} suggests that its semi-classical interpretation in terms of sums over 
space-time backgrounds involves freely acting $\mathbb{Z}_{n_2}$ orbifolds of an $AdS_2 \times S^2$ background \cite{Banerjee:2008ky, Murthy:2009dq, Dabholkar:2011ec, Dabholkar:2014ema}, as we mentioned
in Subsection \ref{sec:sti}.

In Euclidean space-time, a freely acting $\mathbb{Z}_{n_2}$ orbifold background
is described by the following Euclidean line element \cite{Banerjee:2008ky, Murthy:2009dq, Dabholkar:2011ec, Dabholkar:2014ema},
\bea
ds^2 = v_* \left( (r^2 -1) d \theta^2 + \frac{dr^2}{r^2-1} + d \psi^2 + \sin^2 \psi \, d \phi^2\right)
\;,\quad \theta \cong \theta + \frac{2 \pi}{n_2} \;,\quad \phi \cong \phi  + \frac{2\pi}{n_2} ,
\label{orbibh}
\eea
and by gauge potentials $A_{\theta}^I$ 
that acquire a constant real part when orbifolding,
\bea
A_{\theta}^I = - i e^I_* \, (r -1) \, d \theta + {\rm Re} A_{\theta}^I \;.
\eea
Likewise, the dual gauge potentials 
${\tilde A}_{\theta I}$ also  acquire a constant real part when orbifolding.

In the following, we will determine the real part of the gauge potentials that support these orbifolds in the context of $N=2$ theories, by resorting to S-duality and viewing it as a symplectic transformation acting on the symplectic vector that comprises the gauge potentials in these $N=2$ theories. 
We will thereby show that $(\text{Re}A^I_\theta,\text{Re}\tilde{A}_{\theta I})$ depend on the integers $m_1,n_1,j$ introduced in \eqref{set1set2enl}.
Recall that the order of the orbifold is given by $n_2 = - a c \gamma$. Hence, we slightly abuse the notation when denoting these orbifolds as $\mathbb{Z}_{n_2}$ and it is important to keep in mind that the geometry and the supporting gauge potentials depend on the integers $m_1,n_1$ and $j$ in addition to $n_2$.

We obtained 
the saddle point action 
\eqref{expmn}
by extremizing the combination \eqref{expc}
that appears in the exponent of 
\eqref{covqef2}. Following \cite{Banerjee:2008ky, Murthy:2009dq, Dabholkar:2011ec, Dabholkar:2014ema},
we will regard \eqref{expmn}
as the saddle point action of a 
$\mathbb{Z}_{n_2}$ orbifold background.
If we write this action as 
\bea
\frac{\pi}{n_2} \left[ \sqrt{\Delta} + i 
\left( q \cdot {\rm Re} A_{\theta} - p
\cdot {\rm Re} {\tilde A}_{\theta}
\right) \right]\;,
\label{actqA}
\eea
where
\bea
 q \cdot {\rm Re} A_{\theta} - p
\cdot {\rm Re} {\tilde A}_{\theta} = 
 - j \, \ell  - 2 m_1 \, n + 2 n_1\, m  \;,
\label{saddAex}
\eea
then the question is what are the constant real parts of the gauge
potentials $(A_{\theta}, {\tilde A}_{\theta})$ 
that support the orbifold background.

To address this question, we will use an $N=2$ language, label the gauge potentials as
$(A_{\theta}^I, {\tilde A}_{\theta I})$, where $I=0, 1, \dots, n$, and consider a specific $N=2$ model.

\subsection{Gauge potentials and S-duality}

We consider the following $N=2$ model, 
\bea
\label{eq:prepot-tree}
F(Y) = - \frac{Y^1 Y^a \eta_{ab} Y^b}{Y^0} \;,\quad a = 2, \dots, n \,
\eea
where $\eta_{ab} $ is a constant symmetric matrix. For the moment, we work with a type II tree-level prepotential and will discuss the effect of adding loop corrections below. 
We denote the electric-magnetic charges in the type IIA basis by $(q_I, p^I)$,
while
in the heterotic basis they will be denoted by ${\cal Q}_ I$ and ${\cal P}^I$.
Both sets are related as follows,
\begin{equation}
\begin{split}
{\cal Q}_ I =&\; (q_0, p^1, q_a) \; ,\\
{\cal P}^I =&\; (p^0, - q_1, p^a) \;.
\end{split}
\end{equation}
The charge bilinears $(m, n, \ell)$ are expressed as follows in terms of the charges \cite{LopesCardoso:1999fsj},
\begin{equation}
\label{chargebilexp}
\begin{split}
m =&\; {\cal P} ^I \, O_{IJ} \, {\cal P}^J = p^0 q_1 + p^a \eta_{ab} p^b \; , \\
n =&\; {\cal Q} _I \, M^{IJ} \, {\cal Q}_J = - q_0 p^1 + \frac14 q_a \eta^{ab} q_b \; , \\
\ell =&\; {\cal Q}_I \, {\cal P}^I = q_0 p^0 - q_1 p^1 + q_a p^a \; ,
\end{split}
\end{equation}
where we introduced the constant matrices
\bea
M^{IJ} = 
\begin{pmatrix}
0 & - \frac12 & \\
- \frac12 & 0 & \\
0 & & \frac14 \, \eta^{ab}
\end{pmatrix} \;,\quad 
O_{IJ} = 
\begin{pmatrix}
0 &-  \frac12 & \\
- \frac12 & 0 & \\
0 & &  \eta_{ab}
\end{pmatrix} \;.
\eea

In the following, we work in the type IIA basis, and we define the constant real parts of the gauge
potentials $(A_{\theta}^I, {\tilde A}_{\theta I })$ to be
\begin{equation}
\label{backA}
\begin{split}
{\rm Re} A_{\theta}^0 =&\; 2 m_1 p^1 - j p^0  \;, \\
{\rm Re} A_{\theta}^1 =&\; 2 n_1 p^0 + j p^1 \;, \\
{\rm Re} A_{\theta}^a =&\; -\frac12 m_1 \eta^{ab} q_b -  \frac12  j p^a \;,\\
{\rm Re} {\tilde A}_{\theta 0} =&\; 0 \;, \\
{\rm Re} {\tilde A}_{\theta 1} =&\; 0 \;, \\
{\rm Re} {\tilde A}_{\theta a} =&\; -2 n_1 \eta_{ab} p^b + \frac12 j q_a  \; .
\end{split}
\end{equation}
Note that these expressions depend on the integers $(m_1, n_1, j)$, but not on the integers $n_2$ and $m_2$.
One readily verifies that 
\bea
q_ I \, {\rm Re} A_{\theta}^I  - p^I \, {\rm Re} {\tilde A}_{\theta I} 
= -  j \, \ell -  2 m_1 \, n + 2 n_1 \, m  \;,
\eea
in accordance
with the saddle point expression \eqref{saddAex}.

We claim that the gauge potentials $(A_{\theta}^I, {\tilde A}_{\theta I })$  defined by \eqref{backA} transform as a symplectic vector under S-duality.
Let us first consider the action of the T-generator of S-duality on the charges $(q_I, p^I)$,  which is represented by the $SL(2, \mathbb{Z})$ matrix 
\bea
\begin{pmatrix}
1 & s  \\
0 & 1 
\end{pmatrix} \;,\quad s \in \mathbb{Z}, 
\eea
and results in the transformation law (see, for instance, Eq. 3.6 in \cite{Cardoso:2008fr})
\begin{equation}
\begin{split}
&p^0 \rightarrow p^0 \;,\quad p^1 \rightarrow p^1 + s p^0 \;,\quad p^a \rightarrow p^a \; , \\
&q_0 \rightarrow q_0 - s q_1\;,\quad q_1 \rightarrow q_1 \;,\quad q_a \rightarrow q_a - 2 s \eta_{ab} p^b \; .
\end{split}
\end{equation}
The gauge potentials $(A_{\theta}^I, {\tilde A}_{\theta I })$ defined by \eqref{backA} should transform precisely in the same manner as the charges $(q_I, p^I)$, i.e.
\begin{equation}
\label{AbackT}
\begin{split}
&{\rm Re} A_{\theta}^0 \rightarrow {\rm Re} A_{\theta}^0 \;,\quad {\rm Re} A_{\theta}^1 \rightarrow {\rm Re} A_{\theta}^1 + s {\rm Re} A_{\theta}^0 \;,\quad 
{\rm Re} A_{\theta}^a \rightarrow {\rm Re} A_{\theta}^a \; , \\
&{\rm Re} {\tilde A}_{\theta 0} \rightarrow {\rm Re} {\tilde A}_{\theta 0} - s {\rm Re} {\tilde A}_{\theta 1}\;,\quad 
{\rm Re} {\tilde A}_{\theta 1} \rightarrow {\rm Re} {\tilde A}_{\theta 1} \;,\quad {\rm Re} {\tilde A}_{\theta a} \rightarrow {\rm Re} {\tilde A}_{\theta a} - 2 s \eta_{ab} {\rm Re} A_{\theta}^b \; . 
\end{split}
\end{equation}
To verify the transformation behaviour \eqref{AbackT} under the action of the T-generator of S-duality, we need to determine the transformation behaviour
of the quantities that enter in \eqref{AbackT}.
The charge bilinears $(m, n, \ell)$  transforms as follows,
\begin{equation}
\label{Ttrans}
\begin{split}
m \rightarrow&\; m \; , \\
n \rightarrow&\; n - s \, \ell + s^2 \, m \;,\quad s \in \mathbb{Z} \; , \\
\ell \rightarrow&\; \ell - 2 s \, m \; .
\end{split}
\end{equation}
The invariance of the microscopic result \eqref{covqef2} under the transformation \eqref{Ttrans} is implemented by demanding the following transformation
of the integers $(m_1, n_1, m_2, n_2, j)$,
\bea
m_1 \rightarrow m_1 \;,\quad
n_1 \rightarrow n_1 - s j - s^2 m_1 \;,\quad j \rightarrow j + 2 s m_1 \;,\quad
m_2 \rightarrow m_2 \;,\quad
n_2 \rightarrow n_2 \;.
\label{Ttransmn}
\eea
With this at hand, it is straightforward to check that the real parts of the gauge potentials \eqref{backA} indeed transform as in \eqref{AbackT}. Owing to the symplectic product structure, this immediately implies that \eqref{actqA} is invariant under \eqref{Ttrans} and \eqref{Ttransmn}. Note that the transformation \eqref{Ttransmn} leaves the locus \eqref{locpo} invariant when supplemented
by the action\footnote{
\label{foot1}
Under an S-duality transformation
$\begin{pmatrix} a & b\\ c & d \end{pmatrix} \in SL(2, \mathbb{Z})
$, the vector 
$\begin{pmatrix} v \\ \sigma \\
\rho \end{pmatrix}$ transforms as $\begin{pmatrix} v \\ \sigma \\
\rho \end{pmatrix} \rightarrow 
M \, \begin{pmatrix} v \\ \sigma \\
\rho \end{pmatrix}$, where
$ M = 
\begin{pmatrix}
ad + bc & \quad  - b d & \quad  - a c \\
- 2 c d & \quad d^2 &\quad  c^2 \\
- 2 ab &\quad  b^2 & \quad a^2
\end{pmatrix} $.
}
of the T-generator on the triplet
$(\rho, \sigma, v)$.
We will return to the  invariance of the microscopic result \eqref{covqef2} under the action of the T-generator below.

Next, let us consider consider the action of the S-generator of S-duality on the charges $(q_I, p^I)$,  which is represented by the $SL(2, \mathbb{Z})$ matrix 
\bea
\begin{pmatrix}
0 & 1  \\
-1 & 0 
\end{pmatrix} \;,
\eea
and results in (see, for instance, Eq. 3.6 in \cite{Cardoso:2008fr})
\begin{equation}
\begin{split}
&p^0 \rightarrow - p^1 \;,\quad p^1 \rightarrow p^0 \;,\quad p^a \rightarrow \frac12 \eta^{ab} q_b \; , \\
&q_0 \rightarrow - q_1\;,\quad q_1 \rightarrow q_0 \;,\quad q_a \rightarrow - 2  \eta_{ab} p^b \; .
\end{split}
\end{equation}
The gauge potentials $(A_{\theta}^I, {\tilde A}_{\theta I })$ defined by \eqref{backA} should transform precisely in the same manner as the charges $(q_I, p^I)$, i.e.
\begin{equation}
\label{AbackS}
\begin{split}
&{\rm Re} A_{\theta}^0 \rightarrow - {\rm Re} A_{\theta}^1 \;,\quad {\rm Re} A_{\theta}^1 \rightarrow  {\rm Re} A_{\theta}^0 \;,\quad 
{\rm Re} A_{\theta}^a \rightarrow \frac12 \eta^{ab}  {\rm Re} {\tilde A}_{\theta b} \; , \\
&{\rm Re} {\tilde A}_{\theta 0} \rightarrow -  {\rm Re} {\tilde A}_{\theta 1}\;,\quad 
{\rm Re} {\tilde A}_{\theta 1} \rightarrow {\rm Re} {\tilde A}_{\theta 0} \;,\quad {\rm Re} {\tilde A}_{\theta a} \rightarrow  - 2  \eta_{ab} {\rm Re} A_{\theta}^b \; .
\end{split}
\end{equation}
To verify the transformation behaviour \eqref{AbackS} under the action of the S-generator of S-duality, we need to determine the transformation behaviour
of the quantities that enter in \eqref{AbackS}.
The charge bilinears $(m, n, \ell)$  transforms as follows,
\bea
m  \rightarrow  n \;,\quad
n  \rightarrow  m  \;,\quad
\ell \rightarrow-  \ell  .
\label{Strans}
\eea
Demanding that the integers $(m_1, n_1, j)$ transform as 
\bea
m_1 \leftrightarrow - n_1 \;,\quad j \rightarrow - j \;,
\label{m1n1S}
\eea
we establish the transformation behaviour \eqref{AbackS} under the action of the S-generator of S-duality. Observe that if we further supplement \eqref{Strans} with the transformation law
\bea
m_2 \rightarrow m_2 \;,\quad n_2 \rightarrow n_2\;,
\label{m2n2S}
\eea
then the combined transformation \eqref{m1n1S} and \eqref{m2n2S} leaves the locus \eqref{locpo} invariant,
provided also that $\rho \leftrightarrow \sigma$ and $v \rightarrow - v$. The latter correctly implements the action of the S-generator on the triplet
$(\rho, \sigma, v)$, cf. footnote \ref{foot1}.

Thus, we have shown that the real parts of the gauge potentials $(A_{\theta}^I, {\tilde A}_{\theta I })$ given in \eqref{backA} transform as a symplectic vector under S-duality, and that their symplectic product with the charge vector $(q_I, p^I)$ yields the correct invariant expected from the saddle point expression \eqref{saddAex}.

Finally, let us display the expressions for the gauge potentials \eqref{backA} for the case when
$a = 1, c = -1, d =1, b= 0, \Sigma = 0$, in which case the parameters $(m_1, n_1, m_2, n_2, j)$ given in \eqref{set1set2enl} take the values
$(- \delta, 0, 0, \gamma, 1)$. Then,
\begin{equation}
\label{backA-canonical}
\begin{split}
{\rm Re} A_{\theta}^0 =&\; -2 \delta p^1 - p^0  \;, \\
{\rm Re} A_{\theta}^1 =&\; p^1 \;, \\
{\rm Re} A_{\theta}^a =&\; \frac12 \delta \eta^{ab} q_b -  \frac12   p^a \;, \\
{\rm Re} {\tilde A}_{\theta 0} =&\; 0 \;, \\
{\rm Re} {\tilde A}_{\theta 1} =&\; 0 \;, \\
{\rm Re} {\tilde A}_{\theta a} =&\; \frac12 q_a  \;,
\end{split}
\end{equation}
and \eqref{saddAex} is given by
\bea
 q \!\cdot {\rm Re} A_{\theta} - p
\cdot {\rm Re} {\tilde A}_{\theta} = 
 -  \ell  + 2  \delta \, n  \; .
\label{saddAex2}
\eea
Including loop corrections to the $N = 2$ prepotential \eqref{eq:prepot-tree} does not modify the discussion of S-duality above. It will however affect the result of the extremization \eqref{actqA}. If we include a one-loop correction of the form 
\begin{equation}
F(Y) = -\frac{Y^1 Y^a\eta_{ab}Y^b}{Y^0} - \frac{i}{2\pi}\log\eta^{24}\Bigl(\frac{Y^1}{Y^0}\Bigr) \, ,
\end{equation}
then the value of \eqref{actqA} will be modified according to Section \ref{sec:ext-redone}. For the case when $a = 1, c = -1, d =1, b= 0, \Sigma = 0$, the loop corrections to \eqref{actqA} will read 
\begin{equation}
\frac{\pi}{\gamma} \left[ \frac{\sqrt{m+4}}{\sqrt{m}}\sqrt{\Delta} + i 
\left( q \cdot {\rm Re} A_{\theta} - p
\cdot {\rm Re} {\tilde A}_{\theta}
\right) - 2 i \alpha \right]\;.
\end{equation}
Comparing with the extremization result \eqref{r2corrph} that was obtained by taking into account the presence of the $\eta^{24} $ terms in 
\eqref{covqef2}, we see that the saddle point value receives contributions from the symplectic product of the real parts of the gauge potentials with the charge vector \eqref{saddAex2}, and from the loop corrections to the $N=2$ prepotential.

\subsection{Invariance under the T-generator}
\label{sec:T-inv}

Let us now show that the microscopic result \eqref{covqef} is invariant under the T-transformation \eqref{Ttrans} of the charge bilinears $(m, n, \ell)$.
Using \eqref{phg1}, the exponents in \eqref{covqef} are invariant under the combined transformation \eqref{Ttrans} and \eqref{Ttransmn} when supplemented by the transformation
\bea
\tau \rightarrow \tau - s \;,
\label{tau-transf-contour}
\eea
which is the reflection of the T-generator action on the variables $\rho,\sigma,v$ after we have integrated over $\rho$ by residue. The transformation \eqref{tau-transf-contour} also leaves the $\tau$-contour of integration \eqref{t12c} invariant, since the shift
\begin{equation}
	\frac{\ell}{2m} \rightarrow \frac{\ell}{2m}-s
\end{equation}
induced by \eqref{Ttrans} gets compensated by the shift in $\tau_1$. The invariance of the contour arises from our initial choice of a charge dependent contour \eqref{Rccond}, as in 
\cite{Cheng:2007ch}.

Since  \eqref{covqef} involves a sum over the integers in the set \eqref{setP} as well as over $\Sigma$, we also need their transformation properties under the action of the T-generator.
Their transformation properties should be such that, when using the expressions \eqref{set1set2enl}, they result in the transformation \eqref{Ttransmn}.
The following set of transformations achieves this: the Greek entries $\alpha, \beta, \gamma, \delta$ are inert under the action of the T-generator,
while the Latin entries $a, b, c, d$  transform as 
\bea
\begin{pmatrix}
a & b \\
c & d 
\end{pmatrix} \longrightarrow \begin{pmatrix}
a & b \\
c & d 
\end{pmatrix} \begin{pmatrix}
1 \; &  \delta s  \\
0 \; & 1 
\end{pmatrix} =
\begin{pmatrix}
a  \; & b +   a  \delta  s  \\
c  \; & d + c \delta  s
\end{pmatrix} \;,
\label{latinTtransf}
\eea
and $\Sigma$ transforms as 
\bea
\Sigma \rightarrow \Sigma - \beta (a d + b c) s - \beta \delta a c s^2 \;.
\label{sigmaTtransf}
\eea
Using these expressions as well as \eqref{newv}, we obtain for the transformation behaviour of $\rho'_0$ and $\sigma'_0$,
\begin{equation}
\begin{split}
\rho'_0  \rightarrow&\; \rho'_0  + 2 ab \beta s + a^2 \beta  \delta s^2 \;, \\
\sigma'_0 \rightarrow&\;  \sigma'_0 + 2 c d \beta s + c^2 \beta \delta s^2 \:,
\end{split}
\end{equation}
where the shifts are integer valued, so that
\bea
\eta^{24}(\rho'_0) \, \eta^{24}(\sigma'_0) \rightarrow \eta^{24}(\rho'_0) \, \eta^{24}(\sigma'_0) \;.
\eea
Note that $a, c$ and the Greek  entries $\alpha, \beta, \gamma, \delta$ are inert under the action of the T-generator.
Now consider a fixed sector  $(a, c, \alpha, \beta, \gamma, \delta)$: the only independent integers that transform are $b \in \mathbb{Z}$ and $\Sigma \in \mathbb{Z}/|ac| \mathbb{Z}$. For a given $s$, the integer
$b$ transforms by a uniform shift $a \delta s$, and hence we have a bijection from $\mathbb{Z}$ into $\mathbb{Z}$ in a given sector 
$(a, c, \alpha, \beta, \gamma, \delta)$. The transformation law of $\Sigma$ is also an integer shift which also depends on $b$. The prescribed order of summation
\cite{Cardoso:2021gfg}, first $\Sigma$ then $b$, implies that the transformation of $\Sigma$ is 1-to-1 and therefore the sum is invariant.

More generally, we may infer the transformation law of the 9 integers $\alpha, \beta, \gamma, \delta, a, b, c, d,\Sigma$  under general S-duality transformations
from the transformation law of the 5 integers $m_1, n_1, m_2, n_2, j$ under the T-generator and the S-generator 
by `inverting' the relations \eqref{set1set2enl}, as follows.
Recall that  $a >0, c< 0, \gamma > 0$. We use the following gcd rules,
\bea
\gcd (n_2, 0) = n_2 > 0\;,\quad \gcd (-a c , -a) = a >0 \;,\quad \gcd (- a c , c) = - c > 0 \;.
\eea
Since $n_2 \neq 0$, we have  $\gcd (n_2, -m_1) \neq 0$. 
Using the expressions \eqref{set1set2enl} and recalling that $\gcd(\gamma, \delta) = 1$, we infer the relations
\bea
- a c = \gcd (n_2, -m_1) \; ,
\label{acgcd}
\eea
as well as 
\begin{equation}
\label{gdmn}
\begin{split}
\gamma =&\; \frac{n_2}{\gcd (n_2, -m_1) } \; , \\
\delta =&\; -  \frac{m_1}{\gcd (n_2, -m_1) } \; .
\end{split}
\end{equation}
Since $\alpha \in \mathbb{Z}/\gamma \mathbb{Z}$, \eqref{gdmn} uniquely specifies $\alpha$ and $\beta$ 
in terms of $n_2$ and $m_1$ via the relation
$\alpha \delta - \beta \gamma = 1 $.

Using \eqref{acgcd}, and 
recalling the relation $j = 1 + 2 bc$, we obtain
\bea
c = - \gcd \left( \gcd (n_2, -m_1), \frac{j-1}{2} \right) \; ,
\eea
as well as
\bea
a = \frac{\gcd (n_2, -m_1) }{ \gcd \left( \gcd (n_2, -m_1), \frac{j-1}{2} \right) } \;.
\eea
Using the relation \eqref{mnjrel} together with 
$j = 1 + 2 bc = 2 a d -1$, we infer that $m_1 n_1 + m_2 n_2 = a b c d$, and hence
\bea
b d = \frac{1-j^2}{4\gcd (n_2, -m_1)}\;,
\label{latpro}
\eea
from which it follows that
\bea
d = \gcd \left( \frac{1-j^2}{4\gcd (n_2, -m_1)}, \frac{j+1}{2} \right)\;,
\eea
and hence 
\bea
b = \frac{ \frac{1-j^2}{4\gcd (n_2, -m_1)}}{\gcd \left( \frac{1-j^2}{4\gcd (n_2, -m_1)}, \frac{j+1}{2} \right) }\;.
\eea
Finally,  
coming back to \eqref{set1set2enl}, $\Sigma$ can be expressed as
\begin{equation}
\Sigma = -\frac{n_1 + bd\alpha}{\gamma} \; .
\end{equation}
Using the map worked out above, this shows that $\Sigma$ can be written in terms of the 4 integers $n_1,m_1,n_2$ and $j$. 
Thus, the 9 integers $\alpha, \beta, \gamma, \delta, a, b, c, d, \Sigma$ can be expressed purely in terms of $m_1, n_1, n_2, j$. Note that this map does not involve the fifth integer $m_2$. This observation will be relevant for the analysis of the quantum entropy function below.

Let us then verify that the T-transformation \eqref{Ttransmn},
when combined with \eqref{sigmaTtransf},
reproduces the transformation law
\eqref{latinTtransf} and keeps the
Greek entries $\alpha, \beta, \gamma, \delta$ inert.
It follows immediately from \eqref{gdmn}, \eqref{acgcd} and \eqref{latpro} that
\bea
\gamma \rightarrow \gamma \;,\quad \delta \rightarrow \delta  \;,\quad \alpha \rightarrow \alpha  \;,\quad \beta \rightarrow \beta \;,\quad ac \rightarrow ac \;,\quad
b d \rightarrow b d +   j \delta s +  a c \delta^2 s^2 \;.
\eea
Thus, the Greek entries 
 $\alpha, \beta, \gamma, \delta$ are indeed invariant under the T-transformation. Introducing the notation 
\begin{equation}
\label{eq:qqp}
\begin{split}
q =&\; \gcd \left( \gcd (n_2, -m_1), \frac{j-1}{2} \right)\;, \\
q' =&\; \gcd \left( \gcd (n_2, -m_1), \frac{j-1}{2} +   m_1 s \right) \; ,
\end{split}
\end{equation}
 we obtain that under the T-transformation,
\bea
a = \frac{\gcd (n_2, -m_1) }{q} \;\rightarrow\;  a' = \frac{\gcd (n_2, -m_1) }{q'} = a \frac{q}{q'}\;,
\eea
i.e. $a q = a' q'$. From \eqref{eq:qqp}, any prime factor of $q$ divides gcd$(n_2,-m_1)$ and $\frac{j-1}{2}$, and therefore is a prime factor of $q'$. Similarly, every prime factor of $q'$ divides $q$, and
 hence $q = q'$. This shows that $a$ does not transform. It then follows from
 $ac \rightarrow ac $ that also $c$ is invariant.
 
 Next, using that under the T-transformation,
 \bea
 a d = \frac12 \left( j +1 \right) \rightarrow ad +   m_1 s \;,
 \label{adtr}
 \eea
and dividing by $a> 0$, we obtain
\bea
d \rightarrow d +  \frac{m_1}{a} s = d +  c \delta s \;.
\eea
Likewise, using \eqref{adtr}, we infer
\bea
b c = j - a d \rightarrow b c +  m_1 s \;,
\eea
and dividing by $c$, we obtain
\bea
b \rightarrow b +  \frac{m_1}{c} s = d +  a \delta s \;.
\eea
This reproduces the transformation of the Latin entries given in \eqref{latinTtransf}. We also observe that, using $n_1 + \gamma \Sigma = - b d \alpha$ and $m_2 + \delta \Sigma = - b d \beta$, the integer $\Sigma$ can be written as 
\begin{equation}
\Sigma = n_1\beta - m_2\alpha \; ,
\end{equation}
which shows that the T-transformation of $n_1$ induces the transformation given in \eqref{sigmaTtransf}.

We do not test
the invariance of the microscopic expression \eqref{covqef} under the S-generator of the S-duality group, because 
the microscopic expression \eqref{covqef} 
is guaranteed to equal  the $\frac14$ BPS single-centred
degeneracy only for $n\geq m$ \cite{Dabholkar:2012nd}.

\section{The quantum entropy function \label{sec:qef}} 

In this section we relate the quantum entropy function for single-centred four-dimensional $\frac14$ BPS black holes in toroidally compactified string theory
to the microscopic degeneracy formula \eqref{res1}. To do so, we 
set up the quantum entropy function in the $N=2$ formalism, review some of its salient features, and then we specialise to the $N=4$ case.

We consider $N=2$ supergravity theories in four dimensions, describing the coupling of $n_V$ $N=2$ vector multiplets and of $n_H$ $N=2$ hyper multiplets to $N=2$ supergravity
in the presence of Weyl square interactions. For latter use, we introduce 
\bea
\chi = 2(n_V - n_H +1) .
\eea

The quantum entropy function $W$ is defined \cite{Sen:2008yk,Sen:2008vm} as a path integral with a Wilson line insertion in a class
of four-dimensional Euclidean backgrounds $B$. Thus, $W = \sum_B W_B$ where $W_B$ denotes the 
functional integral, with suitable boundary conditions imposed, over all fields in string theory in the space-time background $B$.
The backgrounds in question are backgrounds that 
asymptote to a specific $AdS_2 \times S^2$ solution fixed by the attractor mechanism \cite{Ferrara:1995ih,Ferrara:1996dd,Ferrara:1996um}. The latter is supported by Abelian gauge fields as well as constant scalar fields.
The class of backgrounds we will consider comprises Euclidean $AdS_2 \times S^2$ space-times and orbifolds thereof \cite{Banerjee:2008ky,Murthy:2009dq,Dabholkar:2011ec, Dabholkar:2014ema}.
Further, the fields in the functional integral are required to asymptote to values consistent with the specific attractor $AdS_2 \times S^2$ background in question.

For four-dimensional BPS black holes in the $N=2$ supergravity theories mentioned above,
supersymmetric localization\footnote{Some aspects of the localization framework in supergravity (in particular certain choices of boundary conditions) still deserve further scrutiny, primarily because the path-integral is defined on a non-compact space~\cite{Sen:2023dps}. Assuming such subtleties can be addressed, the result one obtains takes the form given in~\eqref{quantumentro}.} reduces
the $N=2$ quantum entropy function
$W_1(q,p)$ to a finite dimensional integral of
the form \cite{Banerjee:2008ky,Dabholkar:2010uh,Gupta:2012cy,Murthy:2013xpa}
\begin{equation}
W_1 (q,p) = \mathcal{N}\int _{\mathcal{C}} 
d \phi \, \mu(\phi,p) \,  Z_{\rm 1-loop}(\phi)  \; e^{\pi [ 4 \, {\rm Im} F(\phi + i p) -  q \cdot \phi] } \, 
\;,
\label{quantumentro}
\end{equation}
where $q \cdot \phi = q_I \, \phi^I$, with $I = 0, \dots, n_V$. The normalization $\mathcal{N}$ is not fixed by localization but should be determined by either computing the expectation value of the identity operator, i.e. without the Wilson line insertion, or by comparison with the microscopic degeneracies. 
Here, $W_1$ denotes the functional integral over all fields in string theory
in an Euclidean background described by 
an $AdS_2 \times S^2$ space-time, with the fields subjected to the boundary conditions mentioned above.  
The localization manifold is labelled by $n_V + 1$ parameters $\{ \phi^ I \}$.
Note that the integral in \eqref{quantumentro} requires specifying a contour $\mathcal{C}$.
The measure $\mu(\phi)$
arises as a result of the localization procedure implemented
on the field configuration space.
The term $Z_{\rm 1-loop} (\phi)$ 
denotes the one-loop determinant over non-BPS directions orthogonal to the localization locus and receives contributions from bulk modes as well as
large diffeomorphisms \cite{Iliesiu:2022kny}.
 The function $F$ entering in 
\eqref{quantumentro} is the holomorphic function $F(Y, \Upsilon)$ that defines the $N=2$ Wilsonian effective action. It depends on complex scalar fields $Y^I$ and $\Upsilon$. The latter resides in a scalar chiral multiplet associated with the Weyl multiplet. $F(\phi + i p) $ and $F(Y, \Upsilon)$ are related by
\bea
F(\phi + i p) = F\left(Y = \tfrac12 (\phi + ip), \Upsilon = -64\right) \; .
\eea
For later use, we define the quantity
\bea
\label{kpot}
e^{- {\cal K}} = i \left( {\bar Y}^I F_I  - Y^I {\bar F}_ I \right)\;,
\eea
where $F_I = \partial F(Y, \Upsilon) / \partial Y^I$.

The 1-loop determinant 
factor $Z_{\rm 1-loop}$ around localizing backgrounds consisting of $\mathbb{Z}_{n_2}$ orbifolds of $AdS_2 \times S^2$
has been determined very recently in \cite{Iliesiu:2022kny} and shown to receive contributions from 
both bulk modes as well as large diffeomorphisms. For $N=2$ supergravity theories based on $n_V$ vector multiplets and $n_H$ hypermultiplets, 
the contribution to $Z_{\rm 1-loop}$ from the bulk modes is given by 
\bea
Z^{N=2}_{\text{bulk}} = e^{-\frac{n_2\chi}{24}(i\pi -2\log n_2 - 24\zeta'(-1))}\,e^{-(1 - n_2 \, \frac{\chi}{24}) {\cal K}}  \;,
\eea
while the contribution from large diffeomorphisms is 
\bea
Z^{N=2}_{\text{bdry}} = \frac{1}{n_2}\,e^{- {\cal K}}\;.
\eea
Thus, we infer that in $N=2$ supergravity theories, 
$Z_{\rm 1-loop}$ is given by
\bea
Z^{N=2}_{\rm 1-loop} = \frac{1}{n_2}\,e^{-\frac{n_2\chi}{24}(i\pi -2\log n_2 - 24\zeta'(-1))}\,e^{- (2 - n_2 \frac{\chi}{24}) {\cal K}}\;.
\eea
When $n_2=1$, which corresponds to the unorbifolded case, $Z_{\rm 1-loop}$ takes the form 
given earlier in
\cite{Murthy:2015yfa,Gupta:2015gga,Jeon:2018kec}
(up to the pure constant that can be absorbed in the normalization of the path integral).
On the other hand, for $N=4$ supergravity theories with $N_V$ $N=4$ vector multiplets, 
the contribution to $Z_{\rm 1-loop}$ from the bulk modes is given by \cite{Iliesiu:2022kny}
\bea
Z^{N=4}_{\text{bulk}} = e^{{\cal K}}  \;,
\eea
which is independent of the number of $N=4$ vector multiplets, while
the contribution from large diffeomorphisms is the same as in $N=2$ theories and given by
\bea
Z^{N=4}_{\text{bdry}} = \frac{1}{n_2}\,e^{-{\cal K}}  \;.
\eea
Thus, for $N=4$ supergravity theories, 
$Z_{\rm 1-loop}$ is given by 
\bea
Z^{N=4}_{\rm 1-loop} = \frac{1}{n_2} \;.
\label{N4orb}
\eea
Finally, for $N=8$ supergravity, 
$Z_{\rm 1-loop}$ is given by \cite{Iliesiu:2022kny}
\bea
\label{zn8}
Z^{N=8}_{\rm 1-loop} = \frac{1}{n_2}\,e^{4 {\cal K}} \;.
\eea
We note that, if we set $n_2=1$ and if we evaluate ${\cal K}$ on the BPS attractor 
at the two-derivative level, in which case $e^{- {\cal K}}|_* = \frac{A}{4\pi G_N}$,
the exponent in $Z_{\rm 1-loop} $ precisely yields the logarithmic area correction
to the BPS black hole entropy \cite{Banerjee:2011jp,Sen:2012kpz}, 
\bea
{\cal S} = \frac{A}{4 G_N} + \# \log \frac{A}{G_N} \;,
\eea
where
\bea
\label{logc}
\# = \Biggl\{
\begin{matrix}
    -4 \;\;\;\;\;\; & N=8\;, \\
    0 \;\;\;\;\;\; & N=4\;, \\
    2 - \frac{\chi}{24} \;\;\;\;\;\; & N=2\;.
\end{matrix}
\eea

Let us now turn to the measure $\mu$ in \eqref{quantumentro}. A proposal 
for the approximate form of the measure $\mu$ was given in 
\cite{LopesCardoso:2006ugz,Cardoso:2008fr} using arguments based on symplectic covariance. 
A different argument in favor 
of this proposal was given in 
\cite{Cardoso:2019avb} and is related to the logaritmic corrections
\eqref{logc}, as follows.
Let us denote the exponent 
in \eqref{quantumentro} by
\begin{equation}
H(\phi, p, q) =  4 \, {\rm Im} F(\phi + i p) -  q \cdot \phi \; , 
\label{Hexp}
\end{equation}
and consider its extremization,
\begin{equation}
\frac{\partial H(\phi, p, q) }{\partial \phi^I} = 0 \;.
\label{saddleeqH}
\end{equation}
Let us assume that for a given set of black hole charges $(q_I, p^I)$
there exists only one non-inflective 
critical  point $\phi_*$, corresponding
to BPS attractor values such that $H(\phi_*, p, q) > 0$.
Expanding $H(\phi, p, q) $ around the 
critical point $\phi_*$, we obtain
\begin{equation}
\label{Hexpsadd}
\begin{split}
H(\phi, p, q) =&\; H(\phi_*, p, q)  + \frac12 \frac{\partial^2 H}{\partial \phi^I \partial \phi^J}\Big|_{\phi_*} \,
\delta \phi^I \, \delta \phi^J + {\cal O} ( (\delta \phi)^3) \; , \\
=&\; H(\phi_*, p, q)  + \frac12  \frac{\partial^2 (4 \,  {\rm Im} F)}{\partial \phi^I \partial \phi^J}\Big|_{\phi_*} \,
\delta \phi^I \, \delta \phi^J + {\cal O} ( (\delta \phi)^3) \; .
\end{split}
\end{equation}
This equals
\begin{eqnarray}
H(\phi, p, q)
= H(\phi_*, p, q)  + \frac12  \, {\rm Im} F_{IJ}\big|_{\phi_*} \,
\delta \phi^I \, \delta \phi^J  + {\cal O} ( (\delta \phi)^3) \;,
\end{eqnarray}
where $F_{IJ} = \partial^2 F/\partial Y^I \partial Y^J$.
Next, we evaluate 
 \eqref{quantumentro} by taking  $\mu \,  Z_{\rm 1-loop}$ at the attractor point \eqref{saddleeqH} and 
by integrating over the fluctuations
$\delta \phi^I \in \mathbb{C}$
in Gaussian approximation,
\begin{eqnarray}
{W}_1(q,p) \approx
\mathcal{N}\,\Biggl(\frac{\mu \,  \, Z_{\rm 1-loop} }{\sqrt{ |\det {\rm Im} F_{KL}|} }\Biggr)\Bigg|_{\phi_*} \, e^{\pi 
H(\phi_*, p, q)} .
\label{Wapproxsymp}
\end{eqnarray}
For this to reproduce the logarithmic corrections in \eqref{logc}, $\mu$ has
to be proportional to 
$\sqrt{|\det {\rm Im} F_{KL}| }$.

The proposal 
\bea
\mu = \sqrt{|\det {\rm Im} F_{KL}| } \; ,
\label{measmu}
\eea
is based on the $N=2$ Wilsonian action, which is encoded in a holomorphic function $F$. 
However, as also discussed in
\cite{LopesCardoso:2006ugz,Cardoso:2008fr}, one needs to allow for the inclusion of non-holomorphic
terms that are necessary to ensure that quantities such as $H$ in 
\eqref{Wapproxsymp} are invariant under duality symmetries.
A systematic way to incorporate these non-holomorphic terms is provided by 
the framework 
of deformed special geometry \cite{Cardoso:2014kwa}, which is based on a non-holomorphic
function ${\hat F}(Y, {\bar Y}, \Upsilon, {\bar \Upsilon}) = F(Y, \Upsilon) + 2 i \Omega (Y, {\bar Y}, \Upsilon, {\bar \Upsilon})$, where $\Omega$ is a real function that
is homogeneous of second degree. The non-holomorphic terms are encoded in $\Omega$.
The presence of $\Omega$ leads to various modifications of \eqref{measmu}, as discussed 
in \cite{LopesCardoso:2006ugz,Cardoso:2008fr} and reviewed in \cite{Cardoso:2019avb}.
One such modification is that \eqref{measmu} becomes multiplied by $e^{4 \pi \Omega}$. 
Based on these arguments, we will take 
$ \sqrt{|\det {\rm Im} F_{KL}| } \, e^{4 \pi \Omega}$ as a candidate for the measure in the quantum entropy function for 
$\frac14$ BPS black holes
in $N=4$ heterotic string theory below.
Note that the proposal for the measure in 
\cite{LopesCardoso:2006ugz} (cf. (4.6) in that paper)
involves additional terms that we will drop in the $N=4$ context.

Let us now discuss the $N=4$ supergravity theory obtained
from compactifying heterotic string theory on a six-dimensional torus.
The resulting supergravity theory is based on the $N=4$ gravity multiplet as well as on $N_V = 22$ $N=4$ vector multiplets. The $N=4$ gravity multiplet contains 6 Abelian gauge fields as well
as 2 real scalar fields that make up the axio-dilaton, while each of the 22 $N=4$ vector multiplets contains 1 Abelian gauge field as well as 6 real scalar fields. 
There are thus $28 = 6 + 22$ Abelian gauge fields in total. We will use
an $N=2$ description of this $N=4$ supergravity theory and discuss the quantum entropy function for this theory using an $N=2$ language. 
As emphasized in 
\cite{Iliesiu:2022kny}, all the $N=2$ multiplets that arise in the decomposition of the $N=4$
multiplets have to be taken into account when determining 
$Z_{\rm 1-loop}$.
An $N=4$ vector multiplet is decomposed
into one $N=2$ vector multiplet and one
$N=2$ hypermultiplet. In the context of BPS black holes, the complex scalar fields residing in the $N=2$ vector multiplets  are subjected to the attractor mechanism, while those residing in the hypermultiplets are set to
arbitrary constant values. On the localizing manifold, the former give rise to the coordinates $\{\phi^a, a = 2,\ldots, 23\}$, while the latter remain fixed to a constant off-shell and do not affect the localization locus. The $N=4$ gravity multiplet decomposes into the $N=2$ gravity multiplet, two $N=2$
spin-$\frac32$ multiplets and one $N=2$ vector multiplet.  An $N=2$ spin-$\frac32$ multiplet contains 2 Abelian
gauge fields.  We will switch off the charges associated with the 4 Abelian gauge fields residing in the 
$N=2$
spin-$\frac32$ multiplets, that is, we will only retain the electric-magnetic charges 
$(q_I, p^I)$ where $I = 0, \dots,  23,$ associated
to the 24 Abelian gauge fields that reside in the $N=2$ gravity multiplet and in the $N_V + 1 = 23$ $N=2$ vector multiplets
that arise in the above decomposition. Likewise, we will retain the 24 complex scalar fields $Y^I$ that reside
in the $N=2$ vector multiplets and that are subjected to the attractor mechanism.
The associated $N=2$ holomorphic function $F(Y, \Upsilon)$ takes the form
\begin{equation}
\label{F0F}
\begin{split}
F(Y,\Upsilon) =&\; F^{(0)} (Y) + 2 i \, \omega (Y, \Upsilon)\;, \\
F^{(0)} (Y) =&\; - \frac{Y^1 Y^a \eta_{ab} Y^b}{Y^0}  \;,\quad \omega (Y, \Upsilon) = \frac{1}{256 \pi} \Upsilon \, \log \eta^{24} (iS) \;,\quad S = - i \frac{Y^1}{Y^0} \;, 
\end{split}
\end{equation}
where $a = 2, \dots, 23$ and $\eta_{ab}$ is a constant symmetric matrix.
On the other hand, the real function $\Omega$ takes the form
\bea
\label{eq:non-holo}
4 \pi\,\Omega(Y, {\bar Y}) = - \log (S + \bar S)^{12}  \;.
\eea
Evaluating the candidate measure using e.g. (4.11) in \cite{LopesCardoso:2006ugz} gives 
\bea
\label{eq:meas}
\sqrt{|\det {\rm Im} F_{KL}| } \, e^{4 \pi \Omega} = 
\frac{\sqrt{| \det[ - \eta_{ab}]|  } }{ 2 |Y^0|^2 (S + \bar S)^2 } \, 
\sqrt{ e^{- 2  {\cal K}} - 4 | (S + \bar S)^2 D_S\partial_S \Omega|^2
} \;,
\eea
where $D_S\partial_S\Omega = \partial_S^2\Omega + 2(S+\bar{S})^{-1}\partial_S\Omega$. 
In \cite{LopesCardoso:2006ugz}, this candidate measure was derived starting from a symplectic invariant measure and further imposing the magnetic attractor equations to change variables as $dY^I d\bar{F}_I \rightarrow dY^I\sqrt{|\det {\rm Im} F_{KL}| } \, e^{4 \pi \Omega}$. 
However, the choice of this measure is only defined up to a symplectic invariant function. 
In order to match with the microscopic counting, we should choose an appropriate such function 
whose net effect is to produce a measure that
retains only the first term under the square root in \eqref{eq:meas}. Hence, our proposal for the measure $\mu$ in the quantum entropy function 
of $\frac14$ BPS black holes in toroidally compactified heterotic string theory is
\bea
\mu  =   \frac{ 
\sqrt{| \det[ - \eta_{ab}]|} } {2 |Y^0|^2 (S + \bar S)^2} \, e^{- {\cal K} } \; ,
\label{measN4}
\eea
where we are instructed to set
$Y^I = \frac12 (\phi^I + i p^I )$ and $\Upsilon = - 64$. 

Using the $N=4$, $n_2=1$ result $Z_{\rm 1-loop} = 1$ given in \eqref{N4orb},
we obtain
\begin{equation}
W_1 (q,p) = \mathcal{N} \int _{\mathcal{C}} 
d \phi \, {\tilde \mu}_1
\; e^{\pi [ 4 \, {\rm Im} F^{(0)} (\phi + i p) -  q \cdot \phi] } \, 
\;,
\label{quantumentron4}
\end{equation}
where 
\bea
{\tilde \mu }_1 = 
\frac{ 
\sqrt{| \det[ - \eta_{ab}]|} } {2 |Y^0|^2 (S + \bar S)^2} \, \frac{e^{- {\cal K} } }{ \eta^{24}(iS) \, \eta^{24} (i{\bar S})}  \; .
\label{tm1n4}
\eea
Note that the exponent in \eqref{quantumentron4} is given in terms of $F^{(0)}$, so that we regard the $\eta^{24}$ terms as being part of the measure. This point of view will be justified below when we consider the dependence of the integrand in \eqref{quantumentron4} on the axio-dilaton $S$.

The exponent in \eqref{quantumentron4} admits an immediate generalization to the case
when the background is a $\mathbb{Z}_{n_2}$ orbifold of the 
$AdS_2 \times S^2$ space-time described by the Euclidean metric \eqref{orbibh}. On the orbifold, we expect
\bea
 4 \, {\rm Im} F^{(0)} (\phi + i p) -  q \cdot \phi \; \rightarrow \; \frac{1}{n_2} 
 \left[ 4 \, {\rm Im} F^{(0)} (\phi + i p) -  q \cdot \phi + i \left( q \cdot {\rm Re} A_{\theta} - p
\cdot {\rm Re} {\tilde A}_{\theta}
\right) 
 \right],
 \eea
 where $( {\rm Re} A_{\theta}, {\rm Re} {\tilde A}_{\theta})$ denote the constant real
 parts of the gauge potentials $( A_{\theta}, {\tilde A}_{\theta})$ that are induced
 by the orbifold action, as described in Section \ref{sec:orbi}. The orbifold action
 will also affect the one-loop determinant and the measure factor, cf. \eqref{N4orb}. The net effect will be a change of ${\tilde\mu}_1$ in \eqref{quantumentron4} to a new measure
 \bea
 {\tilde \mu }_1 \; \rightarrow \; {\tilde \mu }_{n_2} \; ,
 \eea
 whose form will be inferred from the microscopic expression \eqref{covqef2} shortly. Now recall that a $\mathbb{Z}_{n_2}$ orbifold is supported by gauge potentials \eqref{backA} that
depend on the integers $m_1, n_1, j$. 
The orbifold is therefore labelled by the $4$ integers $m_1,n_1,n_2, j$,
and we will denote its contribution 
to the quantum entropy function by $W_{n_2; m_1,n_1,j}$ with
 \bea
W_{n_2; m_1,n_1,j} (q,p) = (-1)^\ell\,\mathcal{N} \int _{\mathcal{C}} 
d \phi \, {\tilde \mu}_{n_2}
\; e^{\frac{\pi}{n_2} 
 \left[ 4 \, {\rm Im} F^{(0)} (\phi + i p) -  q \cdot \phi + i \left( q \cdot {\rm Re} A_{\theta} - p
\cdot {\rm Re} {\tilde A}_{\theta}
\right) \right] } \;.
\label{quantumentron2}
\eea
Our chosen normalization ensures that the unorbifolded case corresponding to $n_2 =1, m_1 = n_1 = 0, j=1$ reduces to \eqref{quantumentron4}, i.e. $W_{1;0,0,1} = W_1$. 
Once the orbifold contributions are computed, one should sum over all $W_{n_2;m_1,n_1,j}$ to obtain the degeneracies from the quantum entropy function.

We now proceed with the evaluation of $W_{n_2; m_1,n_1,j}$
by integrating
 out the scalar fields
$\phi^a$. We will first focus on 
$W_1$ given in \eqref{quantumentron4}.
Since the factor $e^{- {\cal K}}$ in ${\tilde \mu}_1$ depends on these scalar fields,
the integration cannot be performed exactly. We approximate the result by evaluating
$e^{- {\cal K}}$ on the saddle point values for the $\phi^a$, obtained by extremizing 
the exponent in \eqref{quantumentron4}.

 \subsection{The quantum entropy function on $AdS_2 \times S^2$}

The saddle point values of the $\phi^I$ are obtained by solving the attractor equations
\begin{equation}
4 \, \frac{\partial}{\partial \phi^I} \, 
 {\rm Im}  F^{(0)}(\phi + i p)
  =  q_I \;.
  \label{attracq}
\end{equation}
Solving for $\phi^a$ with $a = 2, \dots, 23$ we obtain the following saddle point values,
\begin{equation}
\phi^a_* = -\frac{1}{S + \bar S} \left(\eta^{ab} q_b +  i (S - \bar S) p^a \right) \;,
\label{attpa}
\end{equation}
where $\eta^{ab} \eta_{bc} = \delta^a_c$. 
This yields the saddle point values 
$Y_*^a = \frac12 \left( \phi_*^a + i p^a \right) $. Hence we obtain
\begin{equation}
\begin{split}
e^{- \cal K}\vert_* =&\; \frac{1}{|Y^0|^2 (S + \bar S)} \left( (p^0)^2 n + (p^1)^2 m + p^0 p^1 \ell \right) \\
&\;+ 2 |Y^0|^2 (S + {\bar S}) 
\left( \frac{\partial_S \omega}{(Y^0)^2 } + \frac{\partial_{\bar S} {\bar \omega}}{({\bar Y}^0)^2 } 
\right)
\;,
\end{split}
\end{equation}
where the charge bilinears are given by \eqref{chargebilexp}.

Next we rewrite $Y^0$ as \cite{LopesCardoso:2006ugz} 
\bea
Y^0 = \frac12 \left( \phi^0 + i p^0 \right) = \frac{p^1 + i {\bar S} p^0}{S + \bar S} \;,
\label{Y0par}
\eea
so that
\begin{equation}
\begin{split}
e^{- \cal K}\vert_* =&\; \frac{ ( S + \bar S) }{|   p^1 + i {\bar S} p^0 |^2} \left( (p^0)^2 n + (p^1)^2 m + p^0 p^1 \ell \right) \\
&\;+ 2 |p^1 + i {\bar S} p^0|^2 (S + {\bar S}) 
\left( \frac{\partial_S \omega}{(p^1 + i {\bar S} p^0)^2 } + \frac{\partial_{\bar S} {\bar \omega}}{(p^1 - i {\bar S} p^0)^2 } 
\right)
\;.
\end{split}
\end{equation}

{From} now on, to simplify the resulting expressions, we set 
\bea
p^0=0 \;.
\eea
We will therefore work
with the following saddle point expressions,
\begin{equation}
\label{muzapp}
\begin{split}
e^{- \cal K}\vert_* =&\;  (S + \bar S) \,  \left( m + 2\,\partial_S \omega + 2\,\partial_{\bar S} {\bar \omega}\right)
\;, \\[1mm]
 {\tilde \mu}_1 \vert_* =&\; \frac{ 
\sqrt{| \det[ - \eta_{ab}]|} } {2 |Y^0|^2 (S + {\bar S})} \, \frac{m + 2\,\partial_S \omega + 2\,\partial_{\bar S} {\bar \omega}
}{ \eta^{24}(iS) \, \eta^{24} (i{\bar S})} \;.
\end{split}
\end{equation}
Then, setting $\tau = i \bar{S} = \tau_1 + i \tau_2$ and changing variables as 
\bea
d\phi^0 \wedge  d \phi^1 = -2i\,\frac{|Y^0|^2 }{\tau_2} \, d\tau \wedge d \bar{\tau}  \;,
\label{d0d1}
\eea
we obtain
\bea 
d\phi^0 \wedge  d \phi^1 
\,  {\tilde \mu}_1 \vert_* 
 = -\frac{i}{2}\,d\tau \wedge d \bar{\tau}  \, \frac{ 
\sqrt{| \det[ - \eta_{ab}]|} } {\tau_2^2} \, \frac{m + E_2(\tau) + E_2 (- \bar \tau)}{ \eta^{24}(\tau) \, \eta^{24} (-{\bar \tau})} \;, 
\eea
where 
\bea
E_2 (\tau) = \frac{1}{2 \pi i } \frac{d}{d \tau} \log \eta^{24} (\tau) .
\eea
Note that the dependence on $|Y^0|^2$ has cancelled out, and hence so has the dependence on $p^1$. Under this change of variables, the contour $\mathcal C$ also changes, but 
for notational simplicity we
will denote the new contour also by $\mathcal C$.

Thus, using the approximate expression \eqref{muzapp} for the measure, we obtain the following approximate expression
for the quantum entropy function contribution $W_1$ in the $N=4$ setting,
\begin{equation}
W_1(q,p) = -\mathcal{N} \, \frac{i}{2} \, \sqrt{| \det[ - \eta_{ab}]|}\int _{\mathcal{C}} \frac{d\tau\wedge d {\bar \tau}}{  \tau^2_2} \; d \phi^a \, \frac{ m + E_2(\tau) + E_2 (- \bar \tau)}{ \eta^{24}(\tau) \, \eta^{24} (-{\bar \tau})} \, e^{\pi [ 4 \, {\rm Im} 
F^{(0)}(\phi + i p) -  q \cdot \phi]  
}
\label{qent1}
\;.
\end{equation}

Next, using the expression for $F^{(0)}$
given in \eqref{F0F}, we
integrate out the $\phi^a$ in \eqref{qent1} ($a=2, \dots, 23$), which
are Gaussian integrals. Each $\phi^a$ integral yields a factor 
\bea
\frac{1}{\sqrt{ \tau_2}},
\eea
 and we obtain
\begin{equation}
W_1(q,p) = 2^{12}\,\mathcal{N} \, \int _{\mathcal{C}} \frac{d\tau \wedge d {\bar \tau}}{(\tau - \bar{\tau})^{13}} \, \frac{m + E_2(\tau) + E_2 (- \bar \tau)}{ \eta^{24}(\tau) \, \eta^{24} (-{\bar \tau})} \,
e^{\frac{\pi}{\tau_2} \left(
n - \ell \tau_1 + m |\tau|^2\right)
}  
\;,
\label{Wtt}
\end{equation}
where the contour $\mathcal C$ now describes a contour in $(\tau_1, \tau_2)$-space.

We now compare \eqref{Wtt} with the microscopic result \eqref{res}. The condition $n_2 = 1$ implies
$a = - c = d = 1, b=0, - \beta= \gamma =1, \alpha = \delta = 0, \Sigma =0$ (cf. \eqref{setP}),
and hence $m_1 = n_1 = m_2 = 0, j=1$. Both expressions agree provided we set $\mathcal{N} = 2^{-12}$, as already observed in 
\cite{Murthy:2015zzy}. 

This agreement is remarkable, for the following reason. The $N=2$ measure \eqref{eq:meas} was obtained in 
\cite{LopesCardoso:2006ugz}
from a symplectic principle and imposing the magnetic attractor equations. We then further approximate 24 of the remaining integrals by setting 22 scalars at their saddle point values, which leads us to the measure \eqref{muzapp} depending only on $S$ and $\bar{S}$. It would be very interesting to understand how such a measure arises from first principles in an $N=4$ setting.

By comparison, the contour $\mathcal{C}$ in \eqref{Wtt} is determined to be the contour ${\hat C}$ which then microscopically specifies a complex integration contour for $(\phi^0,\phi^1)$. This microscopic data has not been determined directly by the supergravity localization procedure, and it would be valuable to do so in the future.
On this contour, $W_1$ can be brought to the form \eqref{covqef2} by adding a total derivative term, as discussed in 
Subsection \ref{sec:td}.
Neglecting the exponentially suppressed contributions that are encoded in $R$, cf. \eqref{covqef},
the resulting expression reads
\bea
\label{kn4}
W_1(q,p) = \frac{1}{2i\pi}\,\frac{1}{2^{12}}\int_{\hat C} \frac{d \tau \wedge d \bar {\tau}}{(\tau - \bar{\tau})^{2}}\left( 
26 + 2 \pi \frac{m |\tau|^2 + n - \ell \tau_1}{\tau_2}
\right) \, 
 e^{\pi [4 {\rm Im}  {\hat F} (\phi + i p) - q \cdot \phi ]\vert_{\phi_*^a}} ,
\eea
where the $\phi^a$ are evaluated on the attractor value \eqref{attpa}, and where we introduced the non-holomorphic 
function \cite{LopesCardoso:2004law}
\bea
{\hat F} (Y, {\bar Y}) = - \frac{Y^1 Y^a \eta_{ab} Y^b}{Y^0} - \frac{3 i}{\pi} \log \left(  
| \eta^2 (\tau) |^2 \, \tau_2 \right) \;,\quad
\tau = \frac{Y^1}{Y^0} .
\label{def-geo}
\eea
This corresponds to a deformation of $F^{(0)}$ by a non-holomorphic term
$\Psi(\tau, {\bar \tau} ) = \log \left(  
| \eta^2 (\tau) |^2 \, \tau_2 \right)$
and is an example of deformed special
geometry \cite{Cardoso:2014kwa}.

The saddle point evaluation of \eqref{kn4} yields the semi-classical entropy
\begin{equation}
\Big[\pi\left(4\,\text{Im}F^{(0)}(\phi + ip) - q\cdot\phi\right) - 12\log\left(|\eta^2(\tau)|^2\tau_2\right)\Big]_{*} \; ,
\end{equation}
which is the Wald entropy corrected by a non-holomorphic term \cite{LopesCardoso:2004law}.

Note that while in the above we assumed that $p^0 =0$
when deriving $W_1$, the microscopic result \eqref{res} is, of course, also valid when $p^0 \neq 0$.

 \subsection{The quantum entropy function on 
 orbifolds of $AdS_2 \times S^2$}

Next, we turn to $W_{n_2; m_1,n_1,j}$ given in \eqref{quantumentron2}. We integrate out the $\phi^a$ in saddle point approximation, as was done for $W_1$. Each $\phi^a$ integral now yields a factor 
\bea
\frac{\sqrt{n_2}}{\sqrt{ \tau_2}}.
\eea
Then, using \eqref{d0d1} as well as \eqref{saddAex}, we obtain (using the normalization $\mathcal{N}$ and contour $\hat{C}$ determined above from $W_1$)
\bea
W_{n_2; m_1,n_1,j} (q,p) = 2i\frac{(-1)^{\ell}\,n_2^{11}}{\sqrt{\text{det}[\eta_{ab}]}} \int _{\hat{C}} \frac{d\tau \wedge d {\bar \tau}}{(\tau - \bar{\tau})^{12}} \, 
|Y^0|^2 \, {\tilde \mu}_{n_2}|_*
\,
e^{\frac{\pi}{n_2}  \left[
n - \ell \tau_1 + m |\tau|^2 +
 2 i \left( - \frac12 j \, \ell - m_1 \, n + n_1 \, m
  \right) \right]
}  
\;,
\label{apwn2}
\eea
where $ {\tilde \mu}_{n_2}|_*$ denotes the measure $ {\tilde \mu}_{n_2}$ evaluated on the saddle point values $\phi^a_*$ given in \eqref{attpa}. We now compare \eqref{apwn2} with the microscopic result \eqref{res}. The comparison shows that the macroscopic result
is more naturally expressed in terms of the  
9 integers $\alpha,\beta,\gamma,\delta$, $a,b,c,d$ and $\Sigma$
that are subject to the constraints displayed in \eqref{setP}, as
\bea
\label{mn4orb}
 {\tilde \mu}_{n_2}|_* = \frac{\gamma^{12}}{n_2^{25}} \, \frac{\sqrt{|\text{det}[-\eta_{ab}]|}}{2 |Y^0|^2 \, (S+\bar{S}) } \, \frac{m + a^2\, E_2 (\rho'_0) + c^2 \,
 E_2 (\sigma'_0)
 }{ \eta^{24}(\rho'_0) \, 
 \eta^{24} (\sigma'_0)} \;,
 \eea
although, as discussed in Section \ref{sec:T-inv}, this can be equivalently formulated in terms of the 4 integers $m_1,n_1,n_2$ and $j$. The expression \eqref{mn4orb} reduces to the one for 
 ${\tilde \mu}_1 \vert_*$ given in \eqref{muzapp}. To account for the factor $\frac{\gamma^{12}}{n_2^{25}}$ in \eqref{mn4orb}, we write it as
 \bea
 \frac{1}{n_2 }  \frac{1}{ (n_2 \, ac)^{12} } \;.
 \eea
 The first factor $\frac{1}{n_2 }$ is accounted for by the expression for $Z_{\rm 1-loop} $ \eqref{N4orb}.
 The second factor can be accounted for by returning to the candidate measure $ \sqrt{|\det {\rm Im} F_{KL}| } \, e^{4 \pi \Omega}$ discussed above
 and replacing it by
 \bea
  \sqrt{\left|\det \left( \frac{ {\rm Im} F_{KL} }{n_2 \, | a c |}  \right) \right| } \; e^{4 \pi \Omega}\;.
 \eea
 This is similar to what was proposed very recently in \cite{Iliesiu:2022kny} in the context of the quantum entropy function for $\frac18$ BPS black holes in $N=8$ supergravity. However, our $N=4$ story is more involved, as the orbifold saddle points are labeled by 4 integers. It is striking that these four numbers are all that is needed to write down the quantum entropy function, including the measure $\tilde{\mu}_{n_2}|_*$. This points to the fact that the orbifolds labeled by $m_1,n_1,n_2$ and $j$ are sufficient to capture the wall-crossing phenomenon from the supergravity path integral, although it is clear we do not have a first principle understanding of the specific combinations of integers that enter \eqref{mn4orb}, especially through the quantities $\rho'_0$ and $\sigma'_0$.

Adding a total derivative, 
and neglecting the exponentially suppressed contributions that are encoded in $R$ (cf. \eqref{covqef}),
we can write \eqref{apwn2} as
\begin{equation}
\label{eq:W-n2-cov}
\begin{split}
W_{n_2;m_1,n_1,j}(q,p) = \frac{e^{i\pi\varphi}}{\gamma}\frac{(-1)^{\ell}}{4\pi(2iac)^{13}}\int_{\hat{C}}\frac{d\tau\wedge d\bar{\tau}}{\tau_2^2}&\left(26 + \frac{2\pi}{n_2}\frac{m|\tau|^2 - \ell \tau_1 + n}{\tau_2}\right) \times \\[1mm]
&e^{\frac{\pi}{n_2}\left[4\text{Im}F^{(0)}(\phi + i p) - q\cdot\phi\right]_{\phi^a_*} - 12\log\left(\eta^2 (\rho'_0)\eta^2 (\sigma'_0)\tau_2\right)} \; ,
\end{split}
\end{equation}
where $\varphi$ is given in \eqref{phg1}. The expression \eqref{eq:W-n2-cov} is well-suited to provide an alternative gravitational description based on 2D Euclidean wormholes. We expand on this in the next section.

\section{Wormhole interpretation \label{wormholes}}

We return to the Euclidean line element \eqref{orbibh} and bring
the two-dimensional part
\bea
ds^2_2 = v_* \left( (r^2 -1) d \theta^2 + \frac{dr^2}{r^2-1} \right)
\;,\quad  \theta \cong \theta + \frac{2 \pi}{n_2} \;,\quad r > 1 \;, 
\label{lads2}
\eea
to the form \eqref{gads2n2}
given below.
This can be 
done as follows \cite{Sen:2011cn}. 
First, by performing the rescaling
$r \rightarrow n_2 \, r$ and $\theta \rightarrow \theta/n_2$ \cite{Dabholkar:2014ema}, the line
element \eqref{lads2} becomes
\bea
ds^2_2 = v_* \left( (r^2 -s^2) d \theta^2 + \frac{dr^2}{r^2-s^2} \right)
\;,\quad  \theta \cong \theta + 2 \pi \;,\quad s = \frac{1}{n_2} \;.
\label{orbat}
\eea
Next, defining 
\bea
\rho = \frac{1}{2s} \log \frac{r + s }{r - s} \;,\quad 
\rho > 0\;,
\eea
we obtain 
\bea
ds^2_2 = v_* \frac{s^2}{\sinh^2 (s \rho)}
\left(  d \theta^2 + d \rho^2 \right) \;,\quad \theta \cong  \theta + 2 \pi \;,\quad
\rho> 0\;.
\eea
Next, using the map \cite{Sen:2011cn} 
\bea
\tan \left( \frac{s}{2} \left( \sigma + i T \right) \right) 
= \tanh \left( \frac{s}{2} \left( \rho + i \theta \right) \right)
\eea
with $- \pi < s\sigma < 0$ and $- \infty < T < \infty$, 
we obtain the line element 
\bea
\label{gads2n2}
ds^2_2 =  v_* \frac{s^2}{\sin^2 (s \sigma) } \left( d T^2 + d \sigma^2 \right) \;,\quad
- \pi < s\sigma < 0 \;,\quad
- \infty < T < \infty \;,\quad s = \frac{1}{n_2} \;.
\eea
When $s=1$,
this line element describes 
a strip of width $\pi$ that represents global $AdS_2$. 

Both line elements \eqref{lads2} and \eqref{gads2n2}, together with a constant dilaton $\Phi_0 = \frac{1}{8\pi G_N^{(2)}}$, are solutions of 2D Euclidean JT gravity. The appearance of this theory can be understood from a dimensional reduction of the 4D theory as follows. As explained in \cite{Sen:2008yk}, the near-horizon geometry \eqref{orbibh} (and its two-dimensional counterpart \eqref{lads2}) relevant for the quantum entropy function is obtained by considering the near-horizon near-extremal limit of a non-extremal 4D black hole. In the near-horizon region and close to extremality, we can consider the reduction of the 4D gravity theory on a two-sphere whose size is given by a linear perturbation $\Phi$ around the extremal value $v_*$. This perturbation plays the role of the dilaton in the effective 2D theory, and the action after dimensional reduction is, to first order in $\Phi$, given by \cite{Navarro-Salas:1999zer}
\begin{equation}
\label{eq:JT}
S_{\rm{JT}} = -\frac{1}{8\pi G_N^{(2)}}\left[\frac12\int_M d^2 x \sqrt{g} R + \int_{\partial M} dt \sqrt{h} K\right]-\frac{1}{2}\int_M d^2 x \sqrt{g}\,\Phi \Bigl(R + \frac{2}{v_*}\Bigr) - \int_{\partial M} dt \sqrt{h}\,\Phi \Bigl(K - \frac{1}{\sqrt{v_*}}\Bigr) \; ,
\end{equation}
where $K$ is the extrinsic curvature of the manifold $M$ and $h$ is the induced metric on the boundary $\partial M$. Evaluating this action on the orbifolded disc geometry \eqref{lads2} gives 
\begin{equation}
I_{\text{JT}} = -\frac{1}{4n_2\,G_N^{(2)}}\; , 
\end{equation}
while the on-shell action of the orbifolded strip with line element \eqref{gads2n2} vanishes. This shows that the Bekenstein-Hawking entropy of the 4D black hole is recovered by the orbifolded disc geometry as 
\begin{equation}
\label{BH-JT}
-I_{\text{JT}} = \frac{2\pi}{n_2}\,\Phi_0 = \frac{\pi v_*}{n_2 \, G_N^{(4)}} \; ,
\end{equation}
where we used the KK relation $1/G_N^{(2)} = \text{vol}(S^2)/G_N^{(4)} = 4\pi v_*/G_N^{(4)}$ to relate the 2D and 4D Newton constants.
The contribution to the path integral from the geometry \eqref{gads2n2} is suppressed since it has vanishing Euler characteristic. Observe that the entropy \eqref{BH-JT} equals the 
leading term in the exponent of the quantum entropy function \eqref{eq:W-n2-cov} evaluated at the extremum, 
\begin{equation}
\frac{\pi}{n_2} [4\text{Im}F^{(0)}(\phi+ i p)-q\cdot\phi]_{*} = \frac{\pi v_*}{n_2 \, G_N^{(4)}} \; .
\end{equation}

The Bekenstein-Hawking entropy above receives higher-derivative corrections. Focusing on the case $n_2=1$ for the moment, we propose to identify the additional contributions to the exponent in \eqref{kn4} as arising from additional fields living on the orbifolded strip \eqref{gads2n2}, namely 24 chiral and 24 anti-chiral massless periodic scalar fields. To motivate this, we periodically identify the strip coordinate
\begin{equation}
\label{eq:period}
T + 2\pi\tau_{2\,*} \cong T \; ,
\end{equation}
and study the effect of adding the matter fields on the geometry \eqref{gads2n2}. Observe that we impose a fixed periodicity given in \eqref{tau} and specified by the 4D attractor equations. This is necessary to compare results in the 2D picture to the extremum of the 4D quantum entropy function.

As discussed in \cite{Garcia-Garcia:2020ttf}, the matter fields backreact on the geometry. The equation $R+\frac{2}{v_*} = 0$ remains unchanged, but the dilaton equation of motion now acquires a source term due to the scalar fields,
\begin{equation}
\nabla_\mu\nabla_\nu\Phi - g_{\mu\nu}\square\Phi + \frac{1}{v_*}\,g_{\mu\nu}\Phi = -\langle T_{\mu\nu} \rangle \; ,
\end{equation}
where the matter stress-tensor $T_{\mu\nu}$ receives a classical and a quantum contribution \cite{Garcia-Garcia:2020ttf}, $\langle T_{\mu\nu} \rangle = T_{\mu\nu}^\text{class} + T_{\mu\nu}^\text{quantum}$. The classical contribution can be obtained by solving the massless wave equation for the scalars. We will restrict ourselves to backreacted configurations which are time-independent, in which case the mode expansion of each bosonic scalar collapses to a constant and ensures that $T_{\mu\nu}^\text{class} = 0$. Note that this corresponds to the absence of imaginary source terms deforming the JT action \cite{Garcia-Garcia:2020ttf}. The quantum contribution to $\langle T_{\mu\nu} \rangle$ can be computed by first going to the cylinder geometry $ds^2_{\text{cyl}} = dT^2 + d\sigma^2$ and then mapping the result to the strip geometry. In doing so, one generically picks up a contribution from the conformal anomaly. In our setting, the two sets of the 24 periodic scalars precisely constitute the degrees of freedom of critical closed bosonic string theory, which ensures the absence of the left and right conformal anomalies. Then the contribution to the stress-tensor is a one-loop effect made possible by the periodic identification \eqref{eq:period}, and we assume that there are no contributions from higher-loop effects. In total, the backreaction of the scalars renders the dilaton non-constant as \cite{Garcia-Garcia:2020ttf}
\begin{equation}
\label{eq:dilaton-profile}
\Phi(\sigma) = - 24\,\mathcal{E}(\tau_{2\,*})\Bigl(1 - \frac{\sigma + \frac{\pi}{2}}{\tan\sigma}\Bigr) 
\;,
\end{equation}
where we have chosen the integration constant so that $\Phi$ is periodic. Here
\begin{equation}
\mathcal{E}(x) = \sum_{n\geq 1}\frac{1}{4\pi\sinh^2(\pi n x)} - \frac{1}{24\pi} \, ,
\end{equation}
is a charge-dependent constant since $\tau_{2\,*} = \frac{\sqrt{\Delta}}{{2m}}$ according to \eqref{tau}. 
The resulting non-constant dilaton background yields a finite on-shell value for the action and can be
interpreted as a 2D Euclidean wormhole solution, as discussed in \cite{Garcia-Garcia:2020ttf}. 
In the semi-classical limit of large $\tau_{2\,*}$, this on-shell action is exponentially suppressed
relative to \eqref{BH-JT}. This semi-classical limit also ensures that $\Phi \ll \Phi_0$ (so long as one stays away from the endpoints of the $\sigma$ interval), which is the regime where the dimensionally reduced 4D gravity theory is well approximated by JT gravity.

The path integral of the scalar fields on the geometry \eqref{gads2n2} with the identification \eqref{eq:period} produces the one-loop partition functions
\begin{equation}
\label{eq:1loop-P}
\frac{1}{\eta^{24}(i\tau_{2\,*})}\,\frac{1}{\eta^{24}(i\tau_{2\,*})} \; ,
\end{equation}
which can be complexified as $\tau_{2\,*} \rightarrow -i\tau_*$ and $\tau_{2\,*} \rightarrow i\bar{\tau}_*$, respectively, to produce a contribution of
\begin{equation}
\eta^{-24}(\tau_*)\,\eta^{-24}(-\bar{\tau}_*) \; ,
\end{equation}
which corresponds precisely to the $\omega$ and $\bar{\omega}$ contributions at the extremum of the 4D quantum entropy function, where $\omega$ is defined in \eqref{F0F}. Our picture therefore suggests that the 2D Euclidean wormholes provide an alternative way to obtain the contribution to the 4D quantum entropy function coming from loop corrections to the supergravity prepotential. Going further, the partition functions \eqref{eq:1loop-P} also include a zero-mode contribution of $1/\sqrt{\tau_{2\,*}}$ for each pair of left- and right-movers, which correctly reproduces the non-holomorphic correction $\log(\tau_{2\,*})^{-12}$ given in \eqref{eq:non-holo}.

Putting the various contributions together, we find that the 2D wormhole picture at $s=1$ accounts for the corrections to the Bekenstein-Hawking entropy,
\begin{equation}
\frac{1}{4G_N^{(2)}} + \log\eta^{-24}(\tau_*) + \log\eta^{-24}(-\bar{\tau}_*) + \log(\tau_{2\,*})^{-12} \; ,
\end{equation}
which nicely match the exponent in the integrand of the 4D quantum entropy function \eqref{kn4} at the extremum. Observe that such a comparison is explicit in the covariant formulation \eqref{kn4} of the localized supergravity path integral rather than in the formulation \eqref{Wtt}, which provides a physical motivation for the addition of the total derivative term that expresses the quantum entropy function in terms of the deformed special geometry given in \eqref{def-geo}.

To discuss the $n_2\neq 1$ interpretation, we should understand how the Bekenstein-Hawking entropy and the saddle point value $\tau_*$ are modified when orbifolding. The former was discussed for general $s=1/n_2$ above and the orbifold simply suppresses the contribution by $n_2$. For the latter, we switch to a 4D point of view where we propose that the two sets of chiral and anti-chiral matter fields are each anchored at different antipodal points of the 2-sphere as a consequence of supersymmetry \cite{Beasley:2006us,Gaiotto:2006ns}. The argument of the Dedekind functions in \eqref{eq:W-n2-cov} consist of $\tau$ and $\bar{\tau}$ dressed by the Greek, Latin and $\Sigma$ integers characterizing the orbifold. The change in the arguments as one moves between antipodal points is given by the operation
\bea
\tau &\rightarrow&\bar {\tau}\;,\quad
\begin{pmatrix}
    a & b\\
    c & d
\end{pmatrix} \rightarrow
\begin{pmatrix}
    0   & -1 \\
    1   &  0
\end{pmatrix}
\begin{pmatrix}
    a & b\\
    c & d
\end{pmatrix} \;,\quad
  \Sigma \rightarrow -\Sigma \; .
\eea
At this stage we do not have a good understanding of the precise dressing of $\tau$ and $\bar{\tau}$ leading to the $\rho'_0$ and $\sigma'_0$ arguments demanded by the microscopic counting in \eqref{newv}. Nevertheless, our wormhole picture leads to  valuable insights into two puzzles. The first is the question of how the 
Euclidean wormholes of \cite{Lin:2022rzw,Lin:2022zxd} contribute to the $N=4$ counting formula. The second has to do with instanton contributions to $N=4$ BPS black hole microstates encoded in the Dedekind functions. What we have seen is that the usual interpretation in terms of marginal decays of two-centred states \cite{Dijkgraaf:1996it,Sen:2007vb} can be equivalently thought of as contributions from 2D wormholes in the $AdS_2$ factor of the near-horizon geometry of the single-centred $\frac14$ BPS black hole. We stress however that the wormholes we have discussed should not be thought of as additional saddle points to the quantum entropy function path integral. Rather, they provide an alternative space-time interpretation of the deformed special geometry \eqref{def-geo} entering the degeneracy formula in the covariant picture. We will come back to this point in the discussion.

We now have two equivalent macroscopic descriptions of $N=4$ BPS black hole microstates, one in terms of the quantum entropy function in 4D $N=4$ supergravity and the other in terms of 2D Euclidean wormholes. Each of these admits a holographic dual and we will now comment on the equivalence between them.

\section{Relating DFF and Liouville type actions in one dimension}
 
For our holographic investigations, we first consider a special class of time-dependent perturbations of the attractor geometry given in 
\cite{Aniceto:2021xhb}.
  These perturbations involve fluctuations of the size of the $S^2$ factor while keeping all other scalars at their attractor values. They have been analyzed in \cite{Castro:2018ffi,Castro:2019vog,Aniceto:2021xhb} who found that their effect can be holographically encoded in a Lorentzian 1D DFF type model \cite{deAlfaro:1976vlx}, whose action is of the form
\bea 
 \int dt \left[ \frac{(\nu')^2}{ \nu} +  \frac{a}{\nu}  \right] \; .
\eea
Here, $\nu(t)$ is the time-dependent perturbation of the size of the $S^2$, $a \in \mathbb{R}$ is defined in terms of the non-extremality parameter as well as the radius of the $AdS_2$ factor in the attractor background, and the action is written in Poincar\'{e} time. Converting to global time coordinates and redefining $dt \rightarrow \alpha (t) \, dt$, the action of the holographic model reads \cite{Gibbons:1998fa}
\bea
S_{\rm DFF}  = \int dt \left[ \frac{(\nu')^2}{ \alpha \, \nu} + \alpha \left( \frac{a}{\nu} + b \, \nu \right) \right] \; ,
\label{dfft}
\eea
where $b\in \mathbb{R}$ is induced by the change to global time coordinate. Setting $\alpha(t) = 1$, taking $a < 0$, $b  < 0$, and substituting $\nu (t) = x^2(t)$ shows that \eqref{dfft} can be brought to the 
form of a DFF action \cite{deAlfaro:1976vlx},
\bea
S_{\rm DFF}  = \int dt \left[ 4 {(x')^2} -  \left( \frac{|a|}{x^2 } + |b| x^2 \right) \right] \;.
\eea
When $a,b > 0$, the parameter $b$ in \eqref{dfft} can be absorbed by scaling $\nu \rightarrow \sqrt{\frac{a}{b}}\,\nu$ and $\alpha \rightarrow \frac{\alpha}{\sqrt{b}}$, which gives
\bea
\label{dfft-scaled}
S_{\rm DFF} = \sqrt{a}\int dt \left[\frac{(\nu')^2}{\alpha\,\nu} + \alpha\left(\frac{1}{\nu} + \nu\right)\right] \; .
\eea

 On the other hand, Liouville type models have recently been shown to arise in the study of holographic duals of
JT gravity \cite{Engelsoy:2016xyb,Mertens:2017mtv} and 
  in the study of certain perturbations of the 2D dilaton in the $AdS_2$ geometry \eqref{gads2n2} underlying the Euclidean wormhole description \cite{Lin:2022rzw,Lin:2022zxd}. 
We now show that the Lorentzian DFF model \eqref{dfft-scaled} can be rewritten as the Liouville model
  shown to underlie the description of 2D Euclidean wormholes \cite{Lin:2022rzw,Lin:2022zxd}.

When considering the time perturbation of the size of the $S^2$, 
one must set
$a > 0$ in \eqref{dfft}, cf. Eq. (3.94) in \cite{Aniceto:2021xhb}. 
Therefore, 
in the following, 
we will take $a>0$ and $b>0$ and simply set $a=b=1$ in view of the scaling freedom mentioned above \eqref{dfft-scaled}. We then relate the DFF type action \eqref{dfft-scaled} to the following Liouville type action
\bea
S_{\rm Liouv} = \int dt \left[ \frac12 (l')^2 + 2 e^{-l} \right] \; ,
\label{liouv}
\eea
where $l(t)$ encodes the perturbation of the 2D dilaton \cite{Lin:2022rzw,Lin:2022zxd}.

To relate $\nu(t)$ to $l(t)$ we compare the terms in the actions \eqref{dfft-scaled} and \eqref{liouv}, and infer the relations
\begin{equation}
\label{ant}
\begin{split}
\alpha \left( \frac{1}{\nu} + \nu \right) =&\; 2 e^{-l} \;, \\
\frac{(\nu')^2}{ \alpha \, \nu} =&\;  \frac12 (l')^2 \;.
\end{split}
\end{equation}
The first equation determines $\alpha(t)$ in terms of $\nu(t)$ and $l(t)$. On the other hand, by combining these two equations we obtain
\bea
(\nu')^2 \left( \frac{1}{\nu^2} + 1 \right) = 4 \left[\left( e^{- \frac12 l }\right)' \right]^2 ,
\label{nl}
\eea
whose solution expresses $\nu(t)$ in terms of $\l(t)$ through
\bea
-\log \left( \frac{1 + \sqrt{1  + \nu^2}}{\nu} \right) + \sqrt{1 + \nu^2} = \pm 2 \, e^{- \frac12 l } \;,
\label{abb0}
\eea
and $\alpha(t)$ is determined by the first equation in \eqref{ant}. Observe that the correspondence between the DFF and Liouville type models involves a non-trivial reparametrization of time.

The equivalence between Liouville and DFF type actions complements the equivalence between the 4D quantum entropy function and the 2D Euclidean wormhole picture we have proposed in the previous section. We see this as a suggestive holographic counterpart of the alternative macroscopic accounting of the $N=4$ BPS black hole microstates based on Euclidean wormholes.

\section{Conclusions}

We have shown that the exact microscopic degeneracy formula extracted from the inverse of the Igusa cusp form lends itself to two rewritings \eqref{res1} and \eqref{expvf}, each endowed with its own
distinct macroscopic interpretation.

The formula \eqref{res1} expresses the single-centred degeneracy as a Rademacher series controlled by a finite number of polar degeneracies corresponding to negative discriminant states. 
The macroscopic
interpretation of \eqref{res1} is in terms of
the quantum entropy function, whose semi-classical limit describes the Wald entropy of a single-centred black hole in the presence of $R^2$ interactions. These interactions are conveniently 
encoded in the $N=2$ Wilsonian holomorphic prepotential $F = F^{(0)} + 2i\omega$, cf. \eqref{quantumentro}. Furthermore, the polar degeneracies are interpreted macroscopically as two-centred $\frac{1}{4}$ BPS bound states together with a continued fraction structure \cite{LopesCardoso:2020pmp}
associated with their decays at walls of marginal stability in the axio-dilaton moduli space. The polar degeneracies come from the degeneracies of the individual 
$\frac{1}{2}$ BPS centres and are generated by the Dedekind functions arising from the instanton correction $\omega$. The full quantum entropy function correctly reproduces the microscopic degeneracies \eqref{res1} upon summing over the individual orbifolded background contributions $W_{n_2;m_1,n_1,j}$ provided we set the measure on the localizing manifold according to \eqref{mn4orb}. This agreement relies on the 1-to-1 map obtained in Section \ref{sec:T-inv} between the four integers $(n_2,m_1,n_1,j)$ characterizing the orbifolds and the eight integers characterizing the set $P$ together with $\Sigma$. It is remarkable that, just as in the case of $\tfrac18$ BPS black holes in $N=8$ supergravity \cite{Iliesiu:2022kny}, such orbifolds are the only gravitational backgrounds needed to recover the microscopic single-centred degeneracies extracted from $1/\Phi_{10}$. This fact points to a vanishing contribution of other types of backgrounds to the quantum entropy function, similar to the situation studied in \cite{Iliesiu:2021are,Heydeman:2020hhw} for supersymmetric indices.

The formula \eqref{expvf} is more naturally intepreted macroscopically in terms of a 2D theory of gravity. The action of the theory is that of JT gravity \eqref{eq:JT} coupled to the field content 
of critical closed bosonic string theory on
Euclidean $AdS_2$ (or an orbifold thereof) with specific boundary conditions. The partition function of the 2D theory receives a classical contribution from the 2D disc geometry, which is precisely the semi-classical entropy obtained in the quantum entropy function formalism at the two-derivative level and encoded in the prepotential $F^{(0)}$. The bosonic string scalars enter at the one-loop level and their effect on the partition function is to generate holomorphic contributions proportional to Dedekind functions as well as non-holomorphic contributions. In the quantum entropy function, these are precisely the instanton correction $\omega$ and the non-holomorphic correction $\Omega$. 
This also prompts us to interpret the instanton and non-holomorphic contributions as being due to 2D Euclidean wormholes. Let us remark that in the case with maximal supersymmetry analyzed in \cite{Iliesiu:2022kny}, such corrections are forbidden by the high amount of supersymmetry. We therefore do not expect a corresponding 2D interpretation to be relevant in the $N=8$ supergravity setting.

This interpretation is also consistent with the associated 1D holographic dual models. The model capturing perturbations of the size of the $S^2$ factor in the four-dimensional black hole near-horizon geometry 
is a Lorentzian DFF type model with action \eqref{dfft-scaled}. This action can be reformulated via an appropriate time reparametrization as a Liouville type model with action \eqref{liouv}, which is holographically dual to perturbations of the 2D dilaton on the global Euclidean $AdS_2$ strip.  These perturbations source 2D Euclidean wormholes.

While the microscopic degeneracies of single-centred $\tfrac14$ BPS black holes are now well understood, our analysis highlights a couple of interesting open problems. From the quantum entropy function perspective, the main issue that remains is to obtain a first principle derivation of the localization measure given in \eqref{mn4orb}. While we have explained how this measure can be obtained from a saddle point approximation of the $N=2$ symplectic invariant measure proposed in \cite{LopesCardoso:2006ugz}, the latter is only defined up to the inclusion of symplectically invariant factors. Symplectic invariance alone therefore cannot suffice in fixing the measure and guarantee an agreement between the quantum entropy function and the microscopic degeneracies. It would be most interesting to understand how \eqref{mn4orb} arises naturally in the $N=4$ setting. To this end, the formulation \eqref{eq:W-n2-cov} of the quantum entropy function may prove useful. From the 2D wormhole perspective, it is crucial to go beyond the identification of the various contributions to the quantum entropy function in a semi-classical approximation, and to clarify the role of wormhole configurations at the quantum level. We hope to report on these questions in the future.

\subsection*{Acknowledgements}
We would like to thank Guillaume Bossard, Kiril Hristov, 
Luca Iliesiu, Emil Martinec, Sameer Murthy, Boris Pioline 
and Ashoke Sen for valuable discussions.
This work was partially
supported by FCT/Portugal through
CAMGSD, IST-ID,
projects UIDB/04459/2020 and UIDP/04459/2020, 
through the LisMath PhD fellowship PD/BD/135527/2018 (MR), by the Riemann Fellowship (AK), and by a public grant as part of the Investissement d'avenir project, reference ANR-11-LABX-0056-LMH, LabEx LMH (VR). AK thanks the Leibniz Universit\"at Hannover (Institute for Theoretical Physics \& the Riemann Center for Geometry and Physics) and the Abdus Salam ICTP for gracious and extended hospitality during key stages of this work, and the Simons Center for Geometry and Physics for hospitality during the final stages of this work. 
MR thanks KU Leuven for its hospitality during part of this work.
GC, SN, VR and MR would also like to thank CERN for its hospitality during the final stages of this work.

\bibliography{N4loc}
\bibliographystyle{utphys}

\end{document}